\documentclass[aps,prd,eqsecnum,nofootinbib,twocolumn,floatfix,superscriptaddress]{revtex4}
\usepackage{graphicx,epsfig}
\usepackage{amsmath,amssymb,amsfonts,bbm} 
\usepackage{dcolumn}
\usepackage{times,mathptm}
\newcommand{\beq}{\begin{equation}}
\newcommand{\eeq}{\end{equation}}
\newcommand{\bea}{\begin{eqnarray}}
\newcommand{\eea}{\end{eqnarray}}

\newcommand{\hbo}{\hbox to 1 true cm {\hfill } } 
\newcommand{\tr}{\hbox{tr}}
\newcommand{\Tr}{\hbox{Tr}}

\newcommand{\co}{ \hbox{ $\cos \alpha $} } 
\newcommand{\si}{ \hbox{ $\sin \alpha $} }
\newcommand{\cco}{ \hbox{ $\cos 2\alpha $} } 
\newcommand{\ssi}{ \hbox{ $\sin 2\alpha $} }
\newcommand{\Id}{ \mathbbm{1} }
\renewcommand{\d}{\delta}

\renewcommand{\b}{\beta}
\renewcommand{\a}{\alpha}

\newcommand{\g}{\gamma}

\newcommand{\m}{\mu}

\newcommand{\s}{\sigma}

\newcommand{\D}{{\Delta}}

\newcommand{\R}{{\cal R}}

\newcommand{\G}{{\large\cal G}}
\newcommand{\U}{{\large\cal U}}
\newcommand{\Z}{{\large\cal Z}}

\renewcommand{\th}{\theta}

\newcommand{\oh}{{\textstyle{\frac{1}{2}}}}

\newcommand{\dg}{\dagger}
\newcommand{\non}{\nonumber}

\newcommand{\rf}[1]{(\ref{#1})}
\newcommand{\ra}{\rightarrow}

\newcommand{\FIGURE}[2][v]{\begin{figure}[#1]#2\end{figure}}
\begin{document}
%
%
\title{Color Screening, Casimir Scaling, and Domain Structure in G(2) and SU(N)
Gauge Theories}

\author{J. Greensite}
\affiliation{The Niels Bohr Institute, Blegdamsvej 17, DK-2100 Copenhagen \O, Denmark}
\affiliation{Physics and Astronomy Dept., San Francisco State
University, San Francisco, CA~94117, USA}

\author{K. Langfeld}
\affiliation{School of Mathematics \& Statistics, 
Plymouth, PL4 8AA, United Kingdom. } 
\affiliation{Institut f\"ur Theoretische Physik, Universit\"at T\"ubingen,
D-72076 T\"ubingen, Germany}
\author{{\v S}. Olejn\'{\i}k}
\affiliation{Institute of Physics, Slovak Academy
of Sciences, SK--845 11 Bratislava, Slovakia}
\author{H. Reinhardt}
\author{T. Tok}
\affiliation{Institut f\"ur Theoretische Physik, Universit\"at T\"ubingen,
D-72076 T\"ubingen, Germany}

\date{\today}
\begin{abstract}

   We argue that screening of higher-representation color charges by gluons implies a domain structure
in the vacuum state of non-abelian gauge theories, with the color magnetic flux in each
domain quantized in units corresponding to the gauge group center.   Casimir scaling of
string tensions at intermediate distances results from random spatial variations in the color magnetic
flux within each domain.  The exceptional G(2) gauge group is an 
example rather than an exception to this picture, although for G(2) there is only one type of vacuum domain, 
corresponding to the single element of the gauge group center.   We present some numerical results 
for G(2) intermediate string tensions and Polyakov lines, as well as results for certain gauge-dependent 
projected quantities.  In this context, we discuss critically the idea
of projecting link variables to a subgroup of the gauge group.  It is argued that such projections 
are useful only when the representation-dependence of the string tension, at some distance scale, is 
given by the representation of the subgroup.
   
\end{abstract}

\maketitle
%
%
\section{Introduction}\label{Introduction}

Over the past decade, a great deal of numerical evidence has been
obtained in favor of the center vortex confinement mechanism (for
reviews, cf.\ refs.\ \cite{review} and \cite{michael}).  According to
this proposal, the asymptotic string tension of a pure non-abelian
gauge theory results from random fluctuations in the number of center
vortices linked to Wilson loops.  It is sometimes claimed that gauge
theory based on the exceptional group G(2) is a counterexample to the
vortex mechanism, in that the theory based on G(2) demonstrates the
possibility of having ``confinement without the center" \cite{Pepe1}.
This claim has two questionable elements.  In the first place, the
asymptotic string tension of G(2) gauge theory is zero, in perfect
accord with the vortex proposal.  No center vortices means no
asymptotic string tension.  From this point of view G(2) gauge theory
is an example of, rather than a counterexample to, the importance of
vortices. The question of whether G(2) gauge theory is nonetheless
confining then depends on what one means by the word ``confinement".
This is a basically a semantic issue, but for the sake of clarity we will
explain our view below. The second point is that G(2), like any group, 
\emph{does} have a center subgroup.  This may seem like a quibble, since the center of G(2) is
trivial, consisting only of the identity element.  We will argue,
however, that this group element has an important role to play, not
only in G(2) but perhaps also in SU(N) gauge theories, in the
classification of non-trival vacuum structures.

   We begin with a discussion of the term
``confinement",   which is used in the literature in several 
inequivalent ways:
\begin{enumerate}
\item Confinement refers to electric flux-tube formation, and a
linear static quark potential.
\item Confinement means the absence of color-electrically 
charged particle states in the spectrum.
\item Confinement is the existence of a mass gap in a
gauge field theory.
\end{enumerate}
We prefer the first of these options. Real QCD, according to this definition, 
confines at intermediate but not at large distance scales, since electric
flux tubes break, the static potential goes flat, and excited hadronic 
states with string-like configurations are only metastable.  For this reason 
we use the term ``temporary confinement" to describe the situation in real 
QCD with light quarks.  Most discussions, however, do not make such a 
distinction, and apply the label ``confinement" to real QCD without any 
qualification.  In that case, to conform to common usage, why not adopt 
the second or third definitions?  Our reason is that this choice would force 
us to also describe many other theories as ``confining" which are not 
normally considered so, and which (unlike real QCD) have nothing at all to do 
with flux tube formation and confining potentials.  

    As one example, let us consider an electric superconductor, or, in its 
relativistic version, the abelian Higgs model.  In this theory, any external electric 
charge is screened immediately by the condensate.  There is a mass gap, and
asymptotic particle states are all electrically neutral.  If definitions (2) and/or (3) 
define confinement, then the abelian Higgs model is surely confining.  But this 
contradicts standard usage in the literature, where confinement of electric charge 
is often identified with a \emph{dual} (i.e.\ magnetic) abelian Higgs model, while
the ordinary abelian Higgs model is identified with confinement of \emph{magnetic}, 
not electric, charge.

    A second example along the same lines is an electrically charged plasma.  In 
response to an external charge,  the plasma rearranges itself to screen out the field 
(and hence the charge) of the probe.  Since all probe charges are neutralized, and all 
Coulombic fields are Debye screened, an electrical plasma could also be taken, according 
to definition (2), as a confining system.

    Our third example is an SU(2) gauge-Higgs theory with a pair of Higgs fields in the 
adjoint representation.  In this case there are two distinct massive phases of the system, 
characterized by an asymptotic string tension which is non-zero in one phase, and zero in 
the other.  We have every right, in this theory, to distinguish between the Confinement 
phase and the Higgs phase.  But definitions (2) and (3) allow no such distinction.  
Both phases, according to these definitions (and according to the 
Fredenhagen-Marcu criterion \cite{FM}) 
are equally confining.\footnote{The Fredenhagen-Marcu criterion, also advocated by
the authors in ref.\ [3] as a confinement criterion, really only tests whether the charge 
of an external source is screened by dynamical matter fields, an effect which also occurs 
in a plasma and an electric superconductor.}  Again, we feel that this is an abuse of the
term ``confinement".   

    Our point is that consistent application of \emph{any} of the three suggested 
definitions does at least some violence to common usage.  According to our 
preferred terminology, i.e.\ option (1),  both real QCD and G(2) gauge theory are
examples of only ``temporary"  confinement.  The other options, in our opinion, 
are worse.  We do not think that electric superconductors, electrical plasmas, and 
gauge systems which are clearly in a Higgs phase, should be described by a term which is 
so often and so generally associated with electric flux tubes and a linear potential.   
 
    Of course, the serious question to be addressed is not whether or not 
to refer to G(2) gauge theory as confining; at the end of the day this is only a matter  of words.
The real challenge which is posed by G(2) gauge theory, in our
opinion, is the fact that this theory has a non-vanishing string
tension for the static quark potential over some finite range of
distances, intermediate between the breakdown of perturbation theory
and the onset of color screening.  Where does that string tension come
from, if not from vortices?  This question has, in fact, a much
broader context.  In connection with SU(N) gauge theories, we have
emphasized many times in the past (beginning with ref.\ \cite{Cas})
that the dependence of the string tension on the color group
representation of the static quarks falls into two general categories,
depending on distance scale:

\begin{enumerate}
\item The Casimir Scaling Region $-$ At intermediate distances, the
  string tension of the static quark potential, for quarks in group
  representation $r$, is approximately proportional to the quadratic
  Casimir $C_{r}$ of the representation, i.e.  
\beq \s_{r} \approx
  \frac{C_{r}}{C_{F}} \s_{F}
     \label{cs}        
     \eeq 
   where $F$ denotes the fundamental, defining representation.
   Thus, e.g., the string tension of quarks in the adjoint
   representation of SU(N) gauge theory is about twice as large as
   that of quarks in the fundamental representation.
\item The Asymptotic Region $-$  Asymptotically, due to color screening by gluons, the string tension
     depends only on the transformation properties of representation $r$ with respect to the center subgroup;
     i.e.\ on the N-ality $k_{r}$ in an SU(N) gauge theory:
     \beq
                  \s_{r} = \s(k_{r})
     \eeq
     where $\s(k)$ is the string tension of the lowest dimensional representation of SU(N) with N-ality $k$.
     These are known as the ``k-string tensions."  The string tension of adjoint representation quarks in
     the asymptotic region is zero.\footnote{Of course, it may be that the k-string tensions are proportional
     to $C_{r_{min}}$, where $r_{min}$ is the lowest dimensional representation of N-ality $k$.  Confusingly,
     this hypothesis is also called ``Casimir scaling" in the literature.  However, the term ``Casimir scaling"
     was originally introduced in ref.\ \cite{Cas} to refer to string tensions in the intermediate region, obeying
     eq.\ \rf{cs}, and we will stick to that original definition here.}
\end{enumerate}
    
We have no reason to think that G(2) gauge theory differs from SU(N)
gauge theories with respect to the intermediate/asymptotic behaviors
of the string tension.  First of all, there is certainly a linear static
quark potential in an intermediate region (as we confirm by numerical
simulations), and we think it likely that the representation
dependence of the string tension at intermediate distances follows an
approximate Casimir scaling law.  Casimir scaling has not been checked
numerically for G(2), but for the purposes of this article we will
assume it to be true, pending future investigation.  
Secondly, all G(2) group representations transform in the same way 
with respect to the center subgroup, i.e.\ trivially, and all must have the same 
asymptotic string tension, namely zero, because of color screening by gluons.
Therefore, the existence of a linear potential in G(2) is the subject
of a broader question, relevant to any gauge group: If center vortices
(or the absence of center vortices) explains the values of the
asymptotic string tension, then what accounts for the linear static
quark potential at intermediate distance scales, where the string
tension depends on the quadratic Casimir of the gauge group, rather
than the representation of the gauge group center?

In the next two sections we will present our answer to this question,
which improves, in some crucial ways, on an old model introduced by
two of the present authors, in collaboration with M.\ Faber
\cite{ccv}.  We begin by asking (section \ref{sec:cscale}) what
features, of a typical gauge field vacuum fluctuation, can possibly
account for both Casimir scaling and color screening.  We are led to
postulate (in section \ref{sec:model}) a kind of domain structure in
the vacuum, with magnetic flux in each domain quantized in units
corresponding to the elements of the center subgroup, including the
identity element.  In section \ref{sec:static} we present some lattice
Monte Carlo results for G(2) gauge theory, which confirm (as expected)
the existence of a string tension over some intermediate distance
range.  In section \ref{sec:dominance} we consider fixing the gauge in
G(2) lattice gauge theory, so as to leave either an SU(3), $Z_{3}$, or
SU(2) subgroup unfixed, together with a projection of the G(2) link
variables to the associated subgroup.  While there are some numerical
successes with this procedure, suggestive of SU(3) or $Z_{3}$ or SU(2)
``dominance" in G(2) gauge theory, in the end we were unable, at least 
thus far, to find any correlation between the projected observables and 
gauge-invariant observables.   We think this failure makes a
point which may also be relevant to, e.g., abelian projection in SU(N)
theories: Projection to a subgroup is unlikely to be of much
significance, unless the group representation-dependence of the string
tension, at some distance scale, depends only on the representation of
the subgroup.  Section \ref{sec:conclusions} contains some concluding
remarks.
 
\section{\label{sec:cscale}Casimir Scaling}

Casimir scaling of string tensions is quite natural in non-abelian
gauge theory.  Consider a planar Wilson loop in group representation
$r$.  Take the loop to lie in, e.g., the $x-y$ plane at $z=t=0$, and
imagine integrating over all links, in the functional integral, which
do not lie in the plane, i.e.
\bea
W_{r}(C) &=& {1\over Z} \int DU_{\m}(x,y,z,t) ~ \chi_{r}[U(C)] e^{-S_{W}}
\non \\
   &=&  \int DU_{x}(x,y,0,0) DU_{y}(x,y,0,0) ~\chi_{r}[U(C)] 
\non \\
  & &  \Bigl\{ \int  [DU_{x} DU_{y}]_{z~\mbox{\small and/or}~t \ne 0} 
\non \\
   & &    \int DU_{z}(x,y,z,t) DU_{t}(x,y,z,t) ~ e^{-S_{W}} \Bigr\}
\non \\
&=& {1\over Z} \int DU_{x}(x,y) DU_{y}(x,y) ~ \chi_{r}[U(C)] 
\non \\
  & & \times \exp\Bigl[-S_{eff}[U_{x},U_{y}]\Bigr]
\label{eq:eff}
\eea
where $\chi_{r}$ is the group character in representation $r$, and
$U(C)$ the loop holonomy.  Then the computation of the planar loop 
in $D=4$ dimensions, with the Wilson action $S_{W}$, is reduced to 
a computation in $D=2$ dimensions, with an effective action $S_{eff}$.   
This effective action can be regarded as the lattice
regulation of some gauge-invariant $D=2$ dimensional continuum action
$S_{eff}^{con}$.  Although $S_{eff}^{con}$ is surely non-local,
non-locality can, in some circumstances, be traded for a derivative expansion, so we have
\beq
       S_{eff}^{con}  =  \int d^{2}x ~ \Bigl( a_{0}\mbox{Tr}[F^{2}] 
 + a_{1}\mbox{Tr}[D_{\m} F D_{\m} F] 
              + a_{2}\mbox{Tr}[(F^{4}] + ... \Bigr)
\label{Seff}
\eeq
where $F=F_{xy}$.  Truncating $S_{eff}^{con}$ to the first term, which ought 
to be dominant at large scales, we obtain
\bea 
      W_{r}(C) &\sim& \frac{1}{Z} \int DA_{x}(x,y) DA_{y}(x,y) ~ \chi_{r}[U(C)] \non \\ 
       & & \qquad \times  \exp\left[-a_{0}\int d^{2}x ~ \mbox{Tr}[F^{2}]  \right]      
\label{dimred}
\eea        
Since this is simply gauge theory in $D=2$ dimensions, the functional
integral can be evaluated analytically, and the result is that the
string tension is proportional to the quadratic Casimir $C_{r}$.  It should
be kept in mind, however, that the validity of the derivative expansion in
\rf{Seff} is an assumption.  There might be some non-local terms in
$S_{eff}$ which are not well represented by such an expansion.

The argument for an effective $D=2$ action can also be phrased in terms of the ground state
wavefunctional $\Psi^{D}_{0}[A]$ of $D+1$ dimensional gauge theory in
temporal gauge.  It was originally suggested in ref.\ \cite{Me1} that
for long-wavelength fluctations, the ground state could be
approximated as
\beq
           \Psi^{D}_{0}[A] \sim \exp\left[-\m\int d^{D}x ~ \mbox{Tr}[F^{2}] \right]     
\label{Psi0}
\eeq   
In that case
\bea
           W_{r}(C) &=& \langle \chi_{r}[U(C)]\rangle^{D=4} 
                      = \langle \Psi^{3}_{0}|\chi_{r}[U(C)]|\Psi^{3}\rangle
\non \\
                        &\sim&  \langle \chi_{r}[U(C)]\rangle^{D=3} 
                      = \langle \Psi^{2}_{0}|\chi_{r}[U(C)]|\Psi^{2}\rangle
\non \\
                        &\sim& \langle \chi_{r}[U(C)]\rangle^{D=2} 
\eea
and again the $D=4$ dimensional calculation has been reduced, in two
steps, to the $D=2$ dimensional theory, where the string tensions are
known to obey the Casimir scaling law.\footnote{Dimensional reduction
from four to two dimensions was advocated independently, on quite
different grounds, by Olesen \cite{Poul1}, who (together with
Ambj{\o}rn and Peterson \cite{Poul2}), noted that this reduction
implies eq.\ \rf{cs}. See also Halpern \cite{Marty} and Mansfield 
\cite{Mansfield}.}  The argument hangs on the validity of the
approximation \rf{Psi0}.  This approximation is supported by
Hamiltonian strong-coupling lattice gauge theory, where the ground
state can be calculated analytically \cite{Me2}, and it has also been
checked numerically \cite{Me3}.  
Karabali, Kim, and Nair \cite{Nair}
have pioneered an approach in which the ground state wavefunctional
$\Psi^{2}_{0}[A]$ of $D=2+1$ dimensional gauge theory can be
calculated analytically in the continuum, as on the lattice, in powers
of $1/g^{2}$, and according to these authors the leading term is again
given by eq.\ \rf{Psi0}.  See also Leigh et al.\ \cite{Leigh} and Freidel
\cite{Freidel}.   

Returning to eq.\ \rf{dimred}, we note that what it really says is
that on a $D=2$ slice of $D=4$ dimensions, the field strength of
vacuum fluctuations is uncorrelated from one point to the next.  This
is because, in $D=2$ dimensions, there is no Bianchi identity
constraining the field strengths, and, by going to an axial gauge, the
integration over gauge potentials $A_{x},~A_{y}$ can be traded for an
integration over the field strength $F_{xy}$.  The absence of
derivatives of $F$ in eq.\ \rf{dimred} means that the field strengths
fluctuate independently from point to point.  This is, of course,
absurd at short distance scales, and results from dropping the
derivative terms in eq.\ \rf{Seff}.  On the other hand, suppose that
$F_{xy}$ has only finite-range correlations, with correlation
length $l$ on the two-dimensional surface slice.  In that case,
dropping the derivative terms may be a reasonable approximation to
$S_{eff}$ at scales larger than $l$, and Casimir scaling is obtained
at large distances.

Regardless of the derivation, the essential point is the following:
Casimir scaling is obtained if field strengths on a 2D slice of the 4D
volume have only finite range correlations.\footnote{In this connection,
see also Shoshi et al.\ \cite{Steffen}.}  Then the color magnetic
fluxes in neighboring regions of area $l^{2}$ are uncorrelated, the
effective long-range theory is D=2 dimensional gauge theory, and
Casimir scaling is the natural consequence.

Once Casimir scaling gets started, the next question is why it
ever ends; i.e.\ why the representation dependence suddenly
switches from Casimir scaling to N-ality dependence.  The conventional
answer is based on energetics: As the flux tube gets longer, at some
stage it is energetically favorable to pair create gluons from the
vacuum.  These gluons bind to each quark, and screen the quark color
charge from representation $r$ to representation $r_{min}$ with the
lowest dimensionality and the same N-ality as $r$.  In particular, if
$r_{min}$ is a singlet, then the flux tube breaks.  Note that this means that
the approximations \rf{dimred} and \rf{Psi0}, which imply Casimir scaling,
must break down at sufficiently large distance scales (at least for
$N_{colors}<\infty$).\footnote{For a treatment of the
long-range Yang-Mills effective action in the context of
strong-coupling lattice gauge theory, cf.\ ref.\ \cite{strong}.}
   
While the energetics reasoning for color screening via gluons is  surely 
correct, when we speak of the ``pair creation" of gluons out the vacuum, we are using
the language of Feynman diagrams, and describing, in terms of particle
excitations, the response of the vacuum to a charged source.  However,
the path-integral itself is a sum over field configurations.  When a
Wilson loop at a given location is evaluated by Monte Carlo methods,
the lattice configurations which are generated stochastically have no
knowledge of the location of the Wilson loop being measured, and no
clue of where gluon pair creation should take place.  The Wilson loop
is just treated as an observable probing typical vacuum fluctuations,
not as a source term in the action.  Nevertheless, the Monte Carlo
evaluation must somehow arrive at the same answer that is deduced, on
the basis of energetics, from the particle picture of string breaking.
This leads us to ask the following question: What feature of a typical
vacuum fluctuation can account for both Casimir scaling of string
tensions at intermediate distances, and center-dependence at large
distances?  What we are asking for is a ``field'' explanation of
screening, to complement the particle picture.

An important observation is that if the asymptotic string tension is
to depend only on N-ality, then the response of a large Wilson loop to
confining vacuum configurations must also depend only on the N-ality
of the loop.  Center vortices are the only known example of vacuum
configurations which have this property.  The confinement scenario
associated with center vortices is very well known: random
fluctuations in the number of vortices topologically linked to a
Wilson loop explain both the existence and the N-ality dependence of
the asymptotic string tension.  Casimir scaling, on the other hand,
seems to require short-range field strength correlations with no
particular reference to the center subgroup.  As Casimir scaling and
N-ality dependence each have straightforward, but different
explanations, the problem is to understand how both of these
properties can emerge, in a lattice Monte Carlo simulation, from the
same set of thermalized lattices.

\section{\label{sec:model}A Model of Vacuum Structure}

   Since Casimir scaling is an intermediate range phenomenon, while N-ality  dependence is asymptotic,
we suggest that the random spatial variations in field strength, required for Casimir scaling, are found in
the interior of center vortices.  These interior variations cannot be entirely random however; they are subject 
to the constraint that the total magnetic flux, as measured by loop holonomies, 
corresponds to an element of the gauge group center.   
   
\begin{figure}[h!]
\centerline{\scalebox{0.15}{\includegraphics{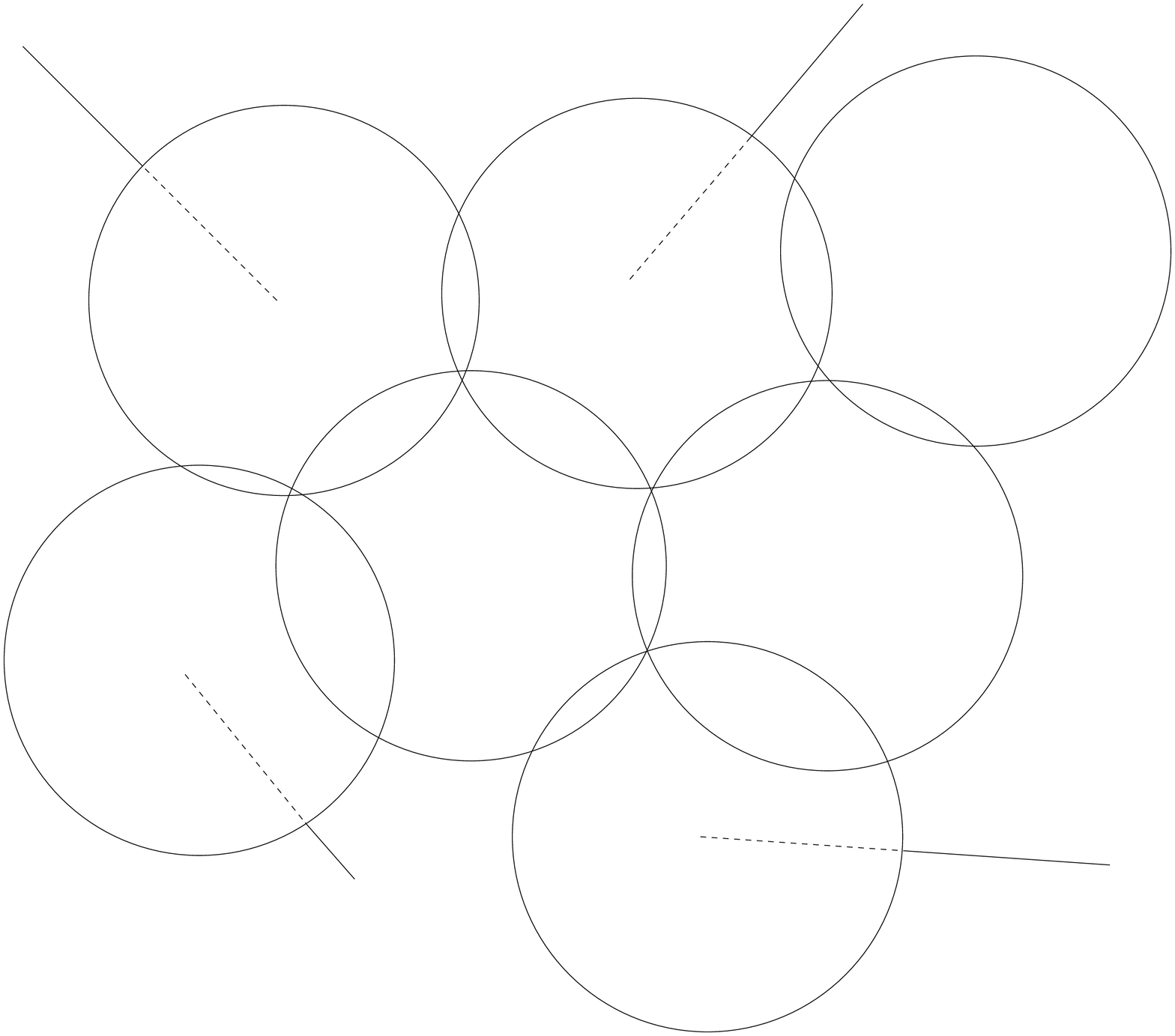}}}
\caption{A 2D slice of the D=4 Yang-Mills vacuum.  Circular regions with (Dirac) lines
correspond to $z=-1$ domains, circular regions without lines denote $z=+1$ domains.}
\label{su2}
\end{figure}

\begin{figure}[h!]
\centerline{\scalebox{0.15}{\includegraphics{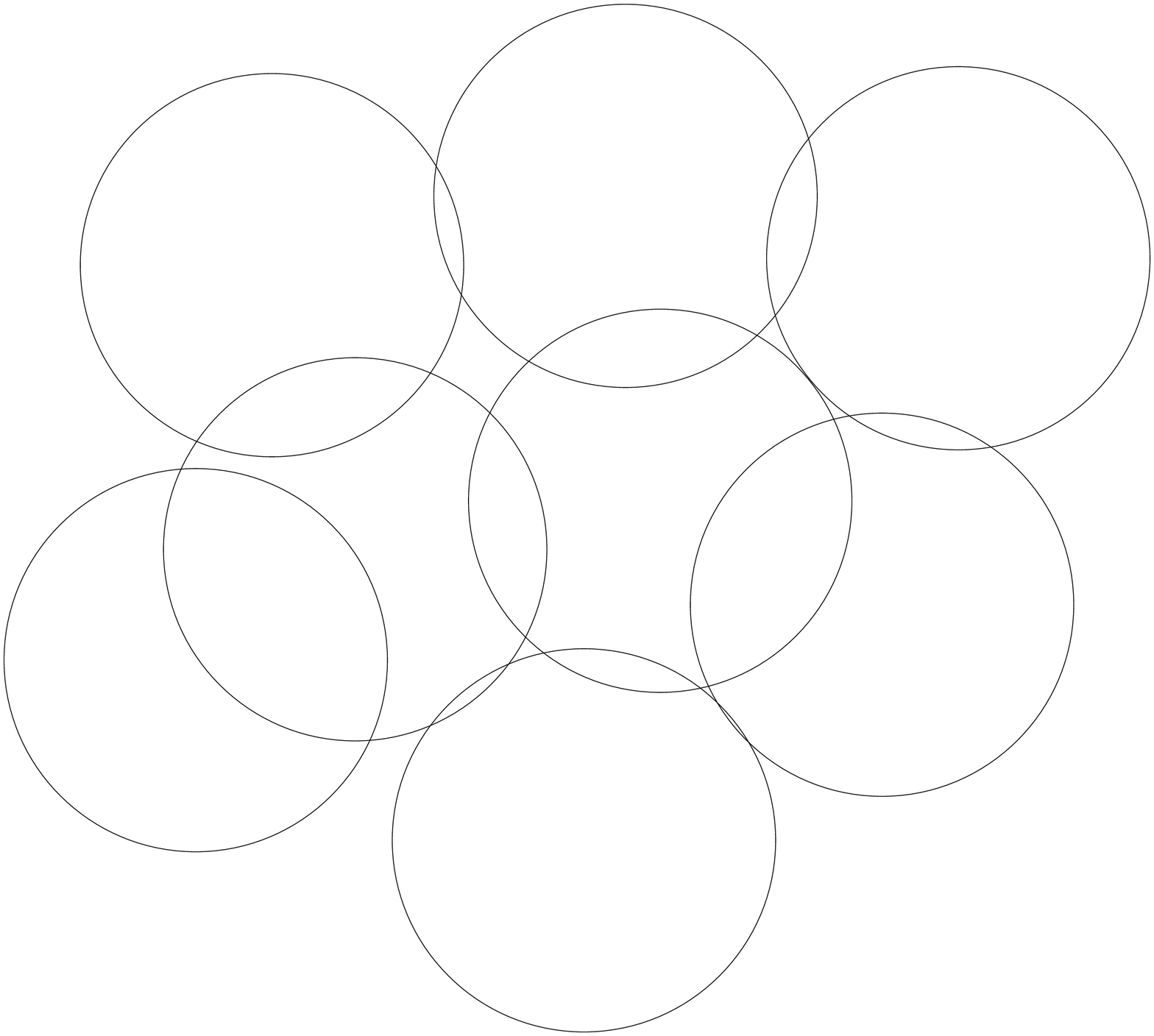}}}
\caption{A 2D slice of the D=4 vacuum of G(2) gauge theory. There is only
one type of domain, corresponding to the single element of the center 
subgroup.}
\label{g2}
\end{figure}

   The picture we have in mind is indicated schematically in Figs.\ \ref{su2} and \ref{g2};
these are cartoons of the vacuum in a 2D slice of the $D=4$ spacetime volume. The first figure 
is an impression of the vacuum in Yang-Mills theory, the second is for G(2) theory.  Each circular
domain is associated with a total flux corresponding to an element of the center subgroup.
The domains may overlap, with the fields in the overlap regions being a superposition of the
fields associated with each overlapping domain.  In the
SU(2) case there are two types of regions, corresponding to the two elements of the center subgroup.
Regions associated with the $Z_{2}$ center element $z=-1$ are 2D slices of the usual thick center vortices,
and the straight lines are projections to the 2D surface of the Dirac 3-volume associated
with each vortex.   But domains corresponding to the $z=1$ center element are also allowed; we see no
reason to forbid them.  Within each domain, the field strength fluctuates 
almost independently in sub-regions of area $l^{2}$, where $l$ is a correlation length for field-strength
fluctuations, subject only to the weak constraint 
that the \emph{total} magnetic flux in each domain corresponds to a center element of the 
gauge group.  In the case of G(2) there is of course only one center element $z=1$ of the trivial $Z_{1}$
subgroup, and no 
Dirac volume, hence the difference in the two cartoons.  This picture is sufficient to explain the
N-ality dependence of loops which are large compared to the domain size, and Casimir
scaling of loops which are small compared to the domain size.

   To support the last statement we return to an old model introduced in ref.\ \cite{ccv}.  
In this model, it is assumed that the effect of a domain (2D cross-section of a vortex)
on a planar Wilson loop holonomy is to multiply the holonomy by a group element
\beq
       G(\a^{n},S) =  S \exp[i\vec{\a^{n}}\cdot \vec{H}] S^\dg
\eeq 
where the $\{H_i\}$ are generators of the Cartan subalgebra, $S$ is a random group element,
$\vec{\a^{n}}$ depends on the location of the domain relative to the loop, and $n$ indicates
the domain type.  If the domain lies entirely within the planar area enclosed by the loop,
then
\beq
            \exp[i\vec{\a^{n}}\cdot \vec{H}]  = z_{n} I
\eeq
where
\beq
          z_{n} = e^{2\pi i n/N} \in Z_{N}
\eeq
and $I$ is the unit matrix.  At the other extreme, if the domain is entirely outside the planar area
enclosed by the loop, then
\beq
            \exp[i\vec{\a^{n}}\cdot \vec{H}]  =   I
\eeq
For a Wilson
loop in representation $r$, the average contribution from a domain will be
\bea
       \G_r[\a^{n}] I_{d_r} &=& \int dS ~ S \exp[i\vec{\a}^{n}\cdot \vec{H}] S^\dg
\non \\
           &=& \frac{1}{d_{r}} \chi_r\Bigl[\exp[i\vec{\a}^{n}\cdot \vec{H}]\Bigr] I_{d_r}
\eea
where $d_r$ is the dimension of representation $r$, and $I_{d_r}$ is the unit matrix.

   Consider, e.g., SU(2) lattice gauge theory, choosing $H=L_3$.  The center subgroup is $Z_{2}$,
and there are two types of domains, corresponding to $z_{0}=1$ and $z_{1}=-1$.  Let $f_{1}$ 
represent the probability that the midpoint of a $z_{1}$ domain is located at any given plaquette
in the plane of the loop, with $f_{0}$ the corresponding probability for a $z_{0}=1$ 
domain.\footnote{In the continuum, $f_{0,1}$ become probability densities to find the midpoints
of domains at any given location.  These should not be confused with the fraction of the lattice covered
by domains, which depends also on domain size.  It is possible that every site on the lattice belongs
to some domain, or to more than one overlapping domains.}
Let us also make the drastic assumption that the probabilities of finding domains of either type centered at any two
plaquettes $x$ and $y$ are independent.  Obviously this ``non-interaction" of domains
is a huge over-simplification, but refinements can come later.  If we make this assumption, then
\bea
   \lefteqn{\langle W_j(C) \rangle} 
\non \\
      &\approx& \prod_x \Bigl\{ (1-f_1-f_0) + f_1\G_j[\a^{1}_C(x)] +
        f_0\G_j[\a^{0}_C(x)] \Bigr\} W_j^{pert}(C)
\non \\
   &=& \exp\left[\sum_x \log\Bigl\{(1-f_1-f_0) + f_1\G_j[\a^{1}_C(x)] \right.
\non \\
        & & \qquad \left. + f_0\G_j[\a^{0}_C(x)] \Bigr\} \right] W_j^{pert}(C)
\eea        
where the product and sum over $x$ runs over all plaquette positions in the plane of the loop $C$, 
and $\a^{n}_C(x)$
depends on the position of the vortex midpoint $x$ 
relative to the location of loop $C$.  The expression $W_j^{pert}(C)$ contains the
short-distance, perturbative contribution to $W_j(C)$; this will just have a perimeter-law falloff.

    To get the static potential, we consider loop $C$ to be a rectangular $R\times T$ contour with
$T \gg R$.  Then the contribution of the domains to the static potential is  
\beq
         V_j(R) = - \sum_{x=\infty}^\infty \log\Bigl\{(1-f_1-f_0) + f_1\G_j[\a^{1}_R(x)] + f_0\G_j[\a^{0}_R(x)]\Bigr\} 
\label{Vr}
\eeq
where $x$ is now the distance from the middle of the vortex to one of the time-like sides of the 
Wilson loop.                                  
The question is what to use as an ansatz for $\a^{n}_R(x)$.  One possibility was suggested in  ref.\
\cite{ccv}, but there the desired properties of the potential in the intermediate region, namely linearity
and Casimir scaling, were approximate at best.  We would like to improve on that old proposal.

    Our improved ansatz is motivated by the idea
that the magnetic flux in the interior of vortex domains fluctuates almost independently, in subregions of
extension $l$,  apart from
the restriction that the total flux results in a center element.  To estimate the effect of this restriction,
consider a set of random numbers $\{F_n\},n=1,2,...,M$,
whose probability distributions are independent apart from the condition
that their sum must equal $2\pi$, i.e.
\beq
        \langle Q[\{F_n\}] \rangle = 
\frac{\int \prod_n dF_n Q[\{F_n\}] e^{-\m \sum_{n=1}^M F_n^2}
                        \d\left(\sum_{n=1}^M F_n - 2\pi \right) }{
                        \int \prod_n dF_n   e^{-\m \sum_{n=1}^M F_n^2}
                        \d\left(\sum_{n=1}^M F_n - 2\pi \right) }
\eeq
Then for a sum of $A<M$ random numbers
\beq
  \left\langle \left( \sum_{n=1}^A F_i \right)^2 \right\rangle = 
\frac{M}{ 2\mu} \left[ \frac{A}{ M} - \frac{A^2}{ M^2} \right]
                         + \left(2\pi \frac{A}{ M}\right)^2
\eeq
If the constraint is, instead, that $\sum_{n=1}^{M} F_{n} = 0$, then the second term on the rhs is dropped.
We then postulate that the $\a_C(x)$ phase likewise consists of a sum of contributions from subregions 
in the domain of area $l^{2}$, which lie in the interior of loop $C$.  It is assumed that these contributions
are quasi-independent in the same sense as the $\{F_n\}$.  Then if the total cross-sectional area of a 
vortex overlapping the minimal area of loop $C$ is $A \gg l^{2}$, we take as our ansatz
\bea
         \Bigl(\a^1_C(x)\Bigr)^{2} &=& \frac{A_v}{ 2\mu} \left[
\frac{A}{ A_v} - \frac{A^2}{ A_v^2} \right]
                         + \left(2\pi \frac{A}{ A_v}\right)^2
\non \\
         \Bigl(\a^0_C(x)\Bigr)^{2} &=& \frac{A'_v}{ 2\mu} \left[
\frac{A}{ A'_v} - \frac{A^2}{ A^{'2}_v} \right]                          
\eea 
where $A_{v},A'_{v}$ are the cross-sectional areas of the $n=1$ and $n=0$ domains, respectively.        
For the sake of simplicity, let us suppose that $A'_{v}=A_{v}$, with $l=1$ lattice spacing, and also imagine that the cross-section 
of a vortex is an $L_v \times L_v$ square. In that case
\bea
V_j(R) &=&  - \sum_{x=-\oh L_v}^{\oh L_v + R} \log\Bigl\{(1-f_1-f_0) + f_1\G_j[\a^{1}_R(x)] 
\non \\
 & & \qquad + f_0\G_j[\a^{0}_R(x)]\Bigr\} 
\eea
Now consider two limits: small loops with $R\ll L_v$, and large loops with $R\gg L_v$.  In the
former case
\bea
       V_j(R) &\approx& - \sum_{x=-\oh L_v + R}^{\oh L_v} \log\Bigl\{(1-f_1-f_0) + f_1\G_j[\a^{1}_R(x)]  
\non \\
           & & \qquad  + f_0\G_j[\a^{0}_R(x)] \Bigr\}
\non \\
             &\approx& -L_v \log\Bigl\{(1-f_1-f_0) + f_1\G_j[\a^{1}_R] 
\non \\
           & & \qquad + f_0\G_j[\a^{0}_R]\Bigr\}
\eea
where 
\bea
      \left(a^1_R\right)^{2} &=&  \frac{L_v^2}{ 2\mu} \left(
\frac{R}{ L_v} - \frac{R^2}{ L_v^2} \right)
                         + \left(2\pi \frac{R}{ L_v}\right)^2   
\non \\
                &\approx& \frac{L_v}{ 2\m} R
\non \\
      \left(a^0_R\right)^{2} &=&  \frac{L_v^2}{ 2\mu} \left(
\frac{R}{ L_v} - \frac{R^2}{ L_v^2} \right)  
\non \\
                &\approx& \frac{L_v }{ 2\m} R
\eea
But for small $\th$
\beq
        \G_j(\th) \approx 1 - \frac{\th^2 }{ 6} j(j+1)
\eeq
so putting it all together, we get for $R\ll L_v$ a linear potential
\beq
           V_j(R) =  \frac{f_1+f_0}{ 6} \frac{L_v^2 }{ 2\m} j(j+1) R 
\eeq
with a string tension
\beq
       \s_j =  \frac{f_1+f_0}{ 6} \frac{L_v^2 }{ 2\m} j(j+1)
\eeq
which is proportional to the SU(2) Casimir.  In the opposite limit, for $R\gg L_v$,
\bea
      \lefteqn{V_j(R)}
\non \\
     &\approx& - \sum_{x=\oh L_v}^{R - \oh L_v} \log\Bigl\{(1-f_1-f_0) + f_1\G_j[\a^{1}_R]
                      + f_0\G_j[\a^{0}_R] \Bigr\}  
\non \\
             &\approx& -R \log\Bigl\{(1-f_{1}) + f_1\G_j[2\pi] \Bigr\}
\eea
where the summation runs over domains which lie entirely within the minimal loop area,  
so that  $\a^{1}_R=2\pi, ~ \a^{0}_{R}=0$.  For those values, $\G_j(\a^{1}_{R}) = -1$ 
for $j=$ half-integer, $\G_j(\a^{1}_{R}) = 1$ 
for $j=$ integer, and $\G_j(\a^{0}_{R}) = 1$  for all $j$.  Then $V_j$ is again a linear potential, 
with asymptotic string tension
\beq
         \s^{asy}_j = \left\{ \begin{array}{cl}
                      -\log(1-2f_{1}) & j = \mbox{half-integer} \cr
                             0    & j = \mbox{integer} \end{array} \right.
\eeq
This has the correct N-ality dependence.  One can adjust the free parameter $\m$ to get a
potential for the $j=\oh$ representation which is approximately linear, with the same slope, at all $R$.
 
   In generalizing to SU(N) we must take into account that there are $N$ types of center domains, 
so that
\bea
   \lefteqn{\langle W_r(C) \rangle}
\non \\
      &\approx& \prod_x \left\{ 1 - \sum_{n=0}^{N-1} f_n\left( 1 - \mbox{Re}
        \G_r[\vec{\a}^{n}_C(x)] \right)\right\} W_r^{pert}(C)
\non \\
   &=& \exp\left[\sum_x \log\left\{  1 - \sum_{n=0}^{N-1} f_n\left( 1 - \mbox{Re}
        \G_r[\vec{\a}_C^{n}(x)] \right)  \right\} \right] W_r^{pert}(C) \non \\
\eea    
where
\beq
         \G_r[\vec{\a}^{n}] = \frac{1}{ d_r} \chi_{r}\left[ \exp[i\vec{\a}^{n} \cdot \vec{H}] \right]
\eeq
and
\beq
       f_n=f_{N-n} ~~~~,~~~~ \G_r[\vec{\a}^{n}] = \G_r^*[\vec{\a}^{N-n}]    
\eeq
For very small loops, which are associated with small $\a$, we have that
\bea
         1 - \G_r[\vec{\a}^{n}] &\approx& \oh \a_i^n \a_j^n 
\frac{1}{ d_r}\mbox{Tr}[H^i H^j]
\non \\
              &=&  \frac{1 }{ 2(N^2 - 1)} \vec{\a}^{n} \cdot \vec{\a}^{n} C_r
\eea
where $C_r$ is the quadratic Casimir for representation $r$.  Once again, if 
$\vec{\a}^{n} \cdot \vec{\a}^{n}$  is proportional to the area of the vortex in the interior of
the loop, for small loops, we end up with a linear potential whose string tension is proportional
to the quadratic Casimir.  For large loops, if the vortex domain lies entirely in the loop
interior, we must have
\beq
         \frac{1}{ d_r} \chi_r\Bigl[\exp[i\vec{\a}^{n} \cdot \vec{H}]\Bigr] = e^{i k n\pi/N} 
\eeq
where $k$ is the N-ality of representation $r$.  Again we obtain a linear potential, 
whose string tension depends only on the N-ality.  It is possible to choose the $f_{n}$ so that
for very large loops the k-string tensions follow either the Sine Law, or are proportional
to the Casimir $C_{r_{min}}$; c.f.\ ref.\ \cite{kstring} for a full discussion of this point.

   Essentially, we are proposing a model of vacuum structure which is much like an instanton or 
monopole or caloron gas picture. The essence of such models is that the functional integral 
over all configurations is approximated, in the infrared, by summation over a special set of configurations, 
and fluctuations around those configurations.  In our case, the special set is center domains.  A center 
domain, whether in SU(N) or G(2) gauge theory, is a planar region and a constraint.  The constraint is that 
color magnetic fluctuations within the planar region add up to a center element, as defined by a loop holonomy 
around the boundary of the domain.   The vacuum fluctuations of a gauge field in a plane is then approximated 
by a sum over domain positions, and an integral over the (constrained) magnetic fluctuations within each 
domain.  This idea can surely be elaborated and perfected; the implementation in this section is only a first, 
and certainly over-simplified, attempt at its realization.  
   
\begin{figure}[h!]
\includegraphics[width=8truecm]{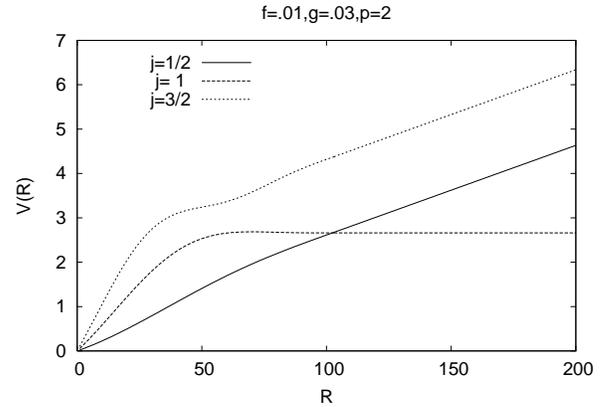}
\caption{The static potential for $j=\oh,1,\frac{3}{2}$ static sources, for
vortex width = 100, in the distance range $R\in [0,200]$.}
\label{cas1}
\end{figure}

\begin{figure}[h!]
\includegraphics[width=8truecm]{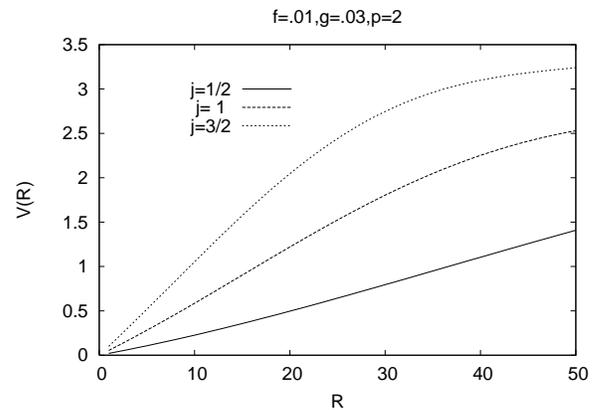}
\caption{Same as Fig.\ \ref{cas1}, in the range $R\in [0,50]$.}
\label{cas2}
\end{figure}

   To illustrate how things may go in SU(2), we choose the following parameters
\begin{itemize}      
\item $L_{v} = 100$ (square cross section)
\item $f_{1}=0.01,~f_{0}=0.03$
\item $L^{2}_{v}/(2\m) = 4$
\end{itemize}
These parameters are chosen as an example; we are not suggesting that they are
necessarily realistic.  We are also taking $l=1$, so the potentials are linear from the
beginning.  Fig.\ \ref{cas1} shows the resulting potentials for $V_{j}(R)$ with 
$j=\oh,1,\frac{3}{ 2}$ in the range
$R\in [0,200]$; one clearly sees the N-ality dependence
at large $R$.   Fig.\ \ref{cas2} shows the same data in a region of smaller $R\le 50$; there
does appear to be an interval whether the potential is both linear and Casimir scaling.  
For fixed $f_{0},f_{1}$ the $\m$ parameter is fine-tuned so that the string tension at 
small and large $R$ are about the same, and the potential for $j=\oh$ is roughly linear over the 
entire range of $R$.\footnote{The string tension shown in Figs.\ \ref{cas1} and \ref{cas2} is linear
even at the smallest values of R because we have taken the field-strength correlation length to
be $l=1$ lattice spacing, and dropped the $W^{pert}(C)$ contribution, which contains the perturbative
contributions to the static potential.}   

   Finally we turn to G(2).  Once again, the logic is that Casimir scaling + screening imply
a domain structure.  The model is much like SU(2) except there is only one type of domain, 
corresponding to the single element of the $Z_{1}$ center subgroup.  The
requirement that the asymptotic string tension vanishes for all representations requires that
$\vec{\a}=0$ when the domain lies entirely in the loop interior, so that   
\beq
     \vec{\a} \cdot \vec{\a}  = \frac{A_v}{ 2\mu} \left[
\frac{A}{ A_v} - \frac{A^2}{ A_v^2} \right]
\eeq    
Together with the fact that Tr${}_r[H^i H^j] \propto \d_{ij} C_r$ in G(2) also,
we obtain a linear potential with Casimir scaling at small $R$, and a vanishing string
tension for large loops, in all representations, at large $R$.

    We note again that Casimir scaling, at the outset of confinement, has never been verified in 
G(2) lattice gauge theory.   However, an initial interval of Casimir scaling (followed by color 
screening) seems to be inescapable in our model, and can be treated as a prediction.  If this 
scaling is not eventually confirmed by lattice simulations in G(2) pure gauge theory, then the 
model is wrong. 

    The last question is how the $z_{0}=1$ domains on a 2D slice extend into the remaining
lattice dimensions, in either G(2) or SU(N) theories.  The $z_{n}\ne 1$ domains, in SU(N) theories,
extend to line-like objects in $D=3$ dimension, or surface-like objects in $D=4$ dimensions, bounding
a Dirac surface or volume, respectively.  These are the usual (thick) center vortices.  We think it
likely that $z_{0}=1$ domains also extend to line-like or surface-like objects in $D=3$ and 4 dimensions.
This opinion is based on the fact that the Euclidean action of vortex solutions, for both $z \ne 1$
\emph{and} $z=1$, has been shown to be stationary at one loop level \cite{Hugo,Diakonov,Bordag}.  While the
stability of these solutions, as well as the validity of perturbation theory at the relevant distance
scales, have not been fully established, the existing results do suggest that the $z=1$ vortices 
are not too different in structure from their $z\ne 1$ cousins.
    
   To summarize:  We have improved on the model introduced in ref.\ \cite{ccv} in two ways.
First, we have motivated an ansatz for the $\a_{C}^{n}(x)$ which gives precise linearity and Casimir
scaling for the static quark potential at intermediate distance scales (``intermediate" begins at
$R=l$).  Secondly,  we have allowed
for the existence of domains corresponding to the center element $z_{0}=1$; this addition permits
a unified treatment of SU(N) and G(2) gauge theories.  The strength of our improved model is that
it gives a simple account of the linearity of the static potential in both the intermediate and asymptotic
distance regimes, with Casimir scaling in the former and N-ality dependence in the latter.  Its main
weakness, in our view, is that there is no explanation for why the string tension of fundamental
representation sources should be the same at intermediate and asymptotic distances.  This equality
can be achieved, but not explained, by fine-tuning one of the parameters in the model.

\section{\label{sec:static}Static Properties of G(2) Gauge Theory}

   In this section we will present some numerical results concerning the intermediate string tension
and Polyakov line values of G(2) lattice gauge theory, determined via Monte Carlo simulations.
    
\subsection{Metropolis Algorithm } 

The partition function is
\beq 
Z \; = \; \int {\cal D} U_\mu \; \exp \left\{ \frac{ \beta }{7} 
\sum _{x, \, \nu > \mu } \tr P_{\mu \nu }(x) \right\} \; ,
\label{eq:a1} 
\eeq 
where $U_\mu (x) $ are $G(2)$ group elements, and 
$P_{\mu \nu }(x) $ is the corresponding plaquette variable. 
In a Metropolis Update step, we monitor the change of action 
under a change of a particular link: 
\beq
U^\prime _\mu (x) \; = \; \Delta (\vec{\alpha } ) \; U_\mu (x) \; , 
\eeq
where $ \Delta (\vec{\alpha } ) \in G(2)$ is given in terms of the 
Euler parameterization (\ref{eq:b1}) as function of the 14 Euler angles,
denoted collectively by 
$\vec{\alpha }$.  These angles are chosen stochastically, in a range  
which is tuned to achieve an acceptance rate of roughly 50\%. 
  
    In order to reduce auto-correlations as well as to reduce 
the number of Monte-Carlo steps, so-called ``micro-canonical reflections" 
are useful: After a certain number of Metropolis steps, the lattice 
configurations is replaced by a configuration of equal action. 
This is done by a sweep through the lattice, and at each link making
the replacement
\beq
U_\mu (x) \; \rightarrow \; U^\prime _\mu (x) \; = \; 
\Delta \; U_\mu (x) \; , \hbo 
S[U_\mu ] \; = \; S[ U^\prime _\mu ] \; . 
\label{eq:m1} 
\eeq
At each link we choose $\Delta = D_k(\alpha )$ for $k = 1 \ldots 7 $. 
Define
\beq 
B_\mu (x) \; = \; \sum _{\nu \not= \mu } P_{\mu \nu }(x) \; , 
\label{eq:stap}
\eeq
where the sum over positive and negative $\nu$ runs over all plaquettes
containing link $(x,\m)$.   Then the replacement link changes the action by
an amount
\beq
\Delta S \; = \; \frac{ \beta }{7}  \Bigl[ \tr  \Delta (\vec{\alpha } ) 
B_\mu (x) \; - \; \tr B_\mu (x) \Bigr] \, . 
\label{eq:a2}
\eeq
the requirement is that at each link
\beq 
\tr \Delta   B_{\mu}(x)  =  
\tr  B_{\mu}(x)  
\eeq
For the choice $\Delta = D_i(\alpha )$, the latter equation can be expressed
in the form 
\beq
a\; \cos \alpha \; + \; b \; \sin \alpha \; + \; c \; = \; 
a \; + \; c \; , 
\eeq
where $a$, $b$ and $c$ are real numbers that depend on $B_{\m}$.
Writing 
\beq
a \; = \; \sqrt{a^2 + b^2 } \; \cos \varphi \; , \hbo 
b \; = \; \sqrt{a^2 + b^2 } \; \sin \varphi \; , 
\eeq 
the constraint becomes 
\beq 
\cos \Bigl( \varphi \, - \, \alpha  \Bigr) \; = \; \cos \varphi \; .
\eeq
which is satisfied by the choice
$$ 
\alpha \; = \; 2 \, \varphi \; . 
$$ 
One also can invent reflections involving the elements 
$D_8 \ldots D_{14}$, but this is not necessary for our purposes.

\begin{figure*}[t]
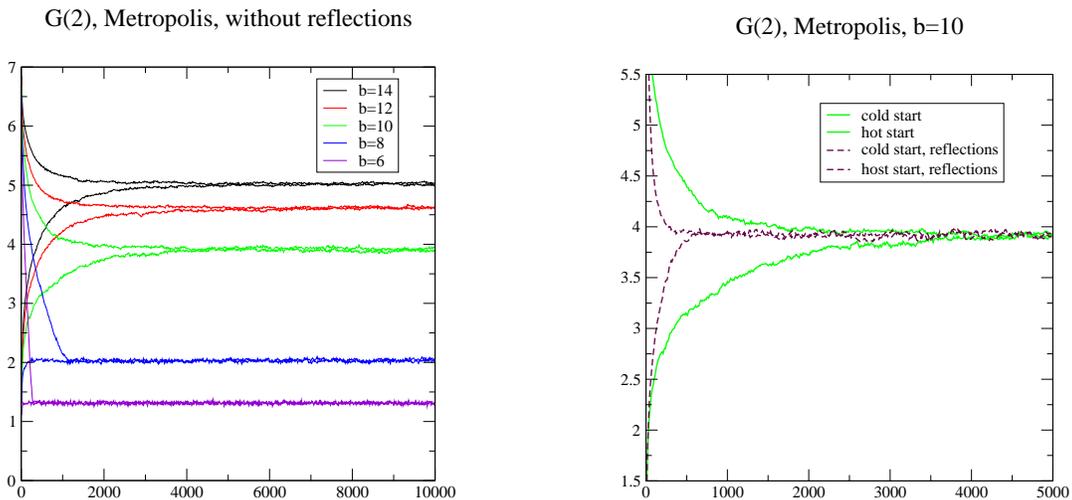

\centerline{  
\epsfxsize=6cm 
\epsffile{therm1.eps} \hspace{2.0cm}
\epsfxsize=6cm 
\epsffile{therm_ref.eps} 
} 
\caption{ The average plaquette values as function of the 
number of Metropolis sweeps for a cold start and a hot start 
for several $\beta $ values: $6^4$ lattice; without (left) and 
with micro-canonical reflections (right).}
\label{fig:1}
\end{figure*}

   To illustrate the usefulness of microcanonical reflections, we display in
Figure~\ref{fig:1} the average plaquette value, as a function of the
number of Monte Carlo sweeps, at several $\b$ values for both cold
and hot starts.   In the strong-to-weak coupling crossover region, at
$\b=10$, some 4000 thermalization sweeps are necessary on a $6^{4}$
lattice volume, when microcanonical relections are not employed.
Figure~\ref{fig:1}, right panel,
shows the improvement if micro-canonical 
reflections are applied. Every ten Metropolis sweeps through the lattice is followed
by a series of four reflection sweeps; the thermalization time is seen to be greatly
reduced.  

\begin{figure}[t]
\includegraphics[width=8truecm]{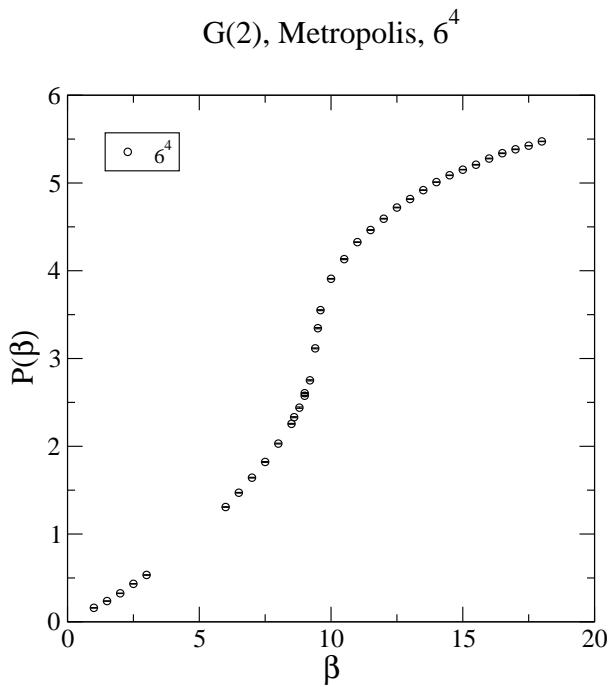} 
\caption{ The plaquette expectation values as function of 
of $\beta $: $6^4$ lattice.}
\label{fig:2}
\end{figure}

    Figure \ref{fig:2} is a plot of the plaquette expectation value
$P (\beta )=\langle \tr P_{\mu \nu }(x) \rangle$ vs.\ $\b$, which
agrees with the result obtained previously in ref.\ \cite{Pepe2}.
A rapid crossover is visible near $\b=10$, but according to
ref.\ \cite{Pepe2} this is not an actual phase transition.  We have 
concentrated our numerical efforts near and just below the crossover, 
in the range $\b \in [9.5,10]$.

\subsection{Static Quark Potential} 

    We are interested in demonstrating the ``temporary confinement"
property of G(2) gauge theory; i.e.\ the linearity of the static quark potential
in some finite distance interval.  The static quark potential can be
calculated from Wilson loops of various sizes, but experience tells us 
that a ``thin''  Wilson line connecting the quark anti-quark sources  
has very little overlap with the ground state in quark anti-quark channel. Spatially 
smeared links (sometimes called ``fat links") are typically used for 
overlap enhancement.  APE 
smearing~\cite{Albanese:1987ds,Teper:1987wt,Perantonis:1988uz} 
adds to particular link a weighted sum of its staples, but the subsequent 
projection onto a SU(N) element usually induces non-analyticities. 
Recently, the so-called ``stout link'' was introduced 
in~\cite{Morningstar:2003gk}; in that approach the link variable
remains in the SU(N) group during smearing. 

All these methods produce time-consuming computer code when
implemented for the G(2) gauge group. For this reason, we will use a
variant of the smearing procedure outlined in~\cite{Langfeld:2003ev},
in which the spatial links for a given time-slice are cooled with
respect to a D=3 dimensional lattice gauge action on the time slice.
Consider a particular spatial link $U_i(x)$, $i = 1 \ldots 3$, and
apply a trial update $U'_{i} = G(\vec{\a})U_{i}$ where \beq
G(\vec{\alpha }) \; = \; \exp \Bigl\{ \alpha _a C^a \Bigr\} \; , \eeq
and the $C^a$, $a=1 \ldots 14$ are the generators of the G(2) algebra.
If we denote by $M$ the sum of space-like plaquette variables
containing the link $U_{i}(x)$ then the choice \beq \alpha _a \; = \;
\epsilon \; \tr \{ C^a \, M \, \} \; , \hbo \epsilon > 0 \; , \eeq
will always lead to an increase in $\tr M$, for sufficiently small
$\epsilon$, if the link $U_{i}(x)$ is replaced by the trial link.  Our
procedure is to start with $\epsilon \approx 1$, and replace links by
trial links if $\tr M$ is increased by the replacement. Proceeding through
the lattice, $\epsilon$ is tuned so that at least 50\% of the
changes are accepted.  One sweep through the lattice is one smearing
step.  As a stopping criterion, one might use the size
of $\Delta S^{(3)} $ or a given number of smearing steps; our present
present results were obtained with a fixed number of $70$ 
smearing steps. %
\begin{figure*}[t]
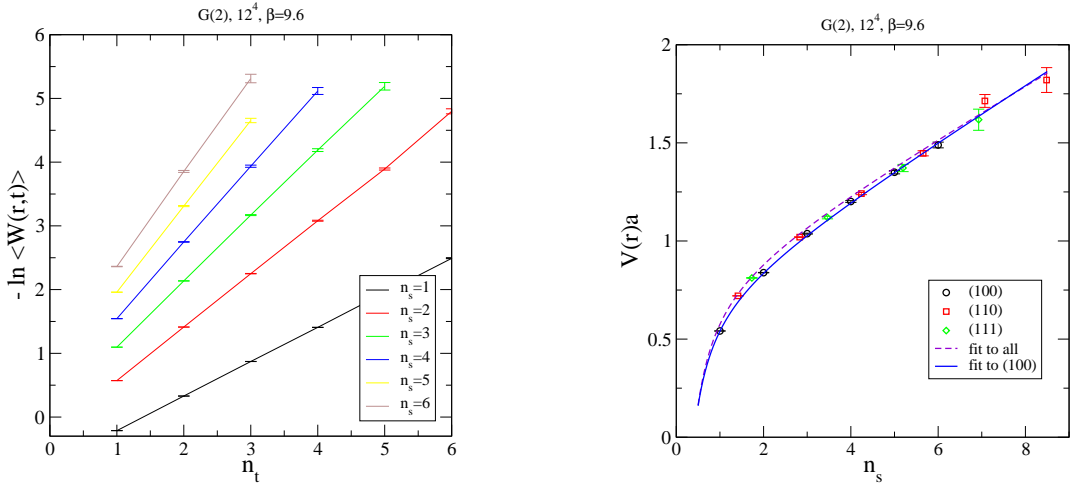
 
\epsfxsize=6cm 
\epsffile{pot_nt.eps} \hspace{2.0cm}
\epsfxsize=6cm 
\epsffile{pot_ns2.eps} 
\caption{ $- \ln \langle W(r,t) \rangle $ as function of $n_t =t/a$ 
for several values of $n_s=r/a$ for a $12^4$ lattice and for $\beta = 
9.6$. The static potential as function of $r$ (right panel).}
\label{fig:4}
\end{figure*}
Let $W(r,t)$ denote the expectation value of a timelike, rectangular
$r\times t$ Wilson loop, where $r=n_{s}a$ and $t=n_{t}a$, and $a$ is
the lattice spacing.  The sides of length $r,~t$ are oriented in
spacelike and time directions, respectively.  The spacelike links are
taken to be the smeared links, while the timelike links are unmodified.
For a fixed value of $r = a n_s$, $V(r)$ is determined from a fit of
$W(r,t)$, computed at various $n_{t}$, to the form
\beq
c \; + \; V(r)a \; n_t \; \approx \; - \ln \; \Bigl\langle W({\cal C} ) 
\Bigr\rangle \; . 
\eeq
Using smeared links, we find that $- \ln \; \langle W({\cal C} )
\rangle $ scales linearly with $n_t$ even for $n_t$ as small as
$n_t=2$ (see figure~\ref{fig:4} for an example at $\b=9.6$). The
corresponding linear fit yields $V(r) a $ including its error bar.
Figure~\ref{fig:4} (right panel) shows $V(r)a$ as function of $r/a$ again
for $\beta = 9.6$. The line between the static quarks (spacelike sides
of the Wilson loops) runs parallel to one of the main axis
(e.g.~$(100)$) of the cubic lattice. In order to check for artifacts
arising from the rotation symmetry breaking, the spacelike sides of
the Wilson loop were also placed along the diagonal axis $(110)$ and
$(111)$.  From Figure~\ref{fig:4} it is clear that at this coupling,
the rotational symmetry breaking effects due to the underlying lattice
structure are quite small.

    The intermediate string tension $\s$ is extracted from a fit of the 
potential $V(r)$ to the form 
\beq
V(r) \; = \; V_0 \; + \; \sigma \, r \; - \; \frac{\alpha }{r} \; , 
\label{eq:p2}
\eeq
A sample fit at $\b=9.6$ is shown in Fig.\ \ref{fig:4} (right panel).
Results from $150$ independent lattice configurations have been 
have been taken into account. Fitting the data from Wilson loops placed 
along the main axis of the lattice ($(100)$-data), we find (with 
$n=r/a$)
$$ 
V(r)a \; = \; 0.713(7) \; - \; \frac{ 0.313(5) }{n} \; + \; 
0.141(2) \, n \; , \; \chi ^2 /\mathrm{dof} \approx 2.1 \; . 
$$
Fitting data from all crystallographic orientations of the Wilson loops, 
we obtain 
$$ 
V(r)a \; = \; 0.79(1) \; - \; \frac{ 0.34(1) }{n} \; + \; 
0.13(2) \, n \; , \; \chi ^2 /\mathrm{dof} \approx 9200 \; . 
$$
The large value for $ \chi ^2 /\mathrm{dof}$ in the latter case 
arises from the facts  that (i) only statistically error bars are 
included in the data and that (ii) the error due to rotational symmetry 
breaking is much larger than the statistical one. Both fits are shown 
in figure~\ref{fig:4} right panel. 

\vskip 0.3cm 
Our numerical findings provide clear evidence of the linearity of the
potential at intermediate distances $r$.  At large distances $r$, one
expects a flattening of the potential due to string breaking effects.
From experience with SU(2) and SU(3) lattice theories, we do not
expect to observe string breaking from measurements of modest-sized
Wilson loops.  String-breaking should be observable in sufficiently
large Wilson loops, but this would require the rather more
sophisticated noise reduction methods employed, e.g., in ref.\
\cite{dFK}.
 
\begin{figure*}[t]
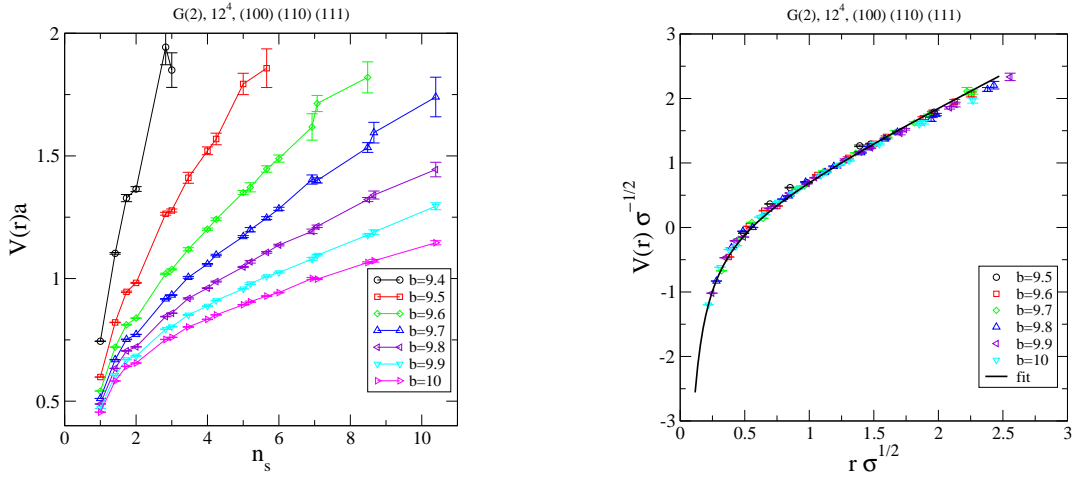
 
\epsfxsize=6cm 
\epsffile{pot_ns_all.eps} \hspace{2.0cm}
\epsfxsize=6cm 
\epsffile{pot_all.eps} 
\caption{ The static potential in units of the lattice 
spacing (left panel) and in units of the intermediate) 
string tension (right panel). }
\label{fig:5}
\end{figure*}
\begin{table*}[t!]
\begin{center}
\begin{tabular}{lcccccc} \hline
$\beta $ & $9.5$ & $9.6$ & $9.7$ & $9.8 $ & $9.9$ & 
$10$  \\ \hline 
$\sigma a^2$ &  $0.24(2)$ & $0.14(1) $ & $0.102(3)$ & $0.079(1)$ 
& $0.060(2)$  & $0.047(1)$ \\ \hline 
$\alpha $ &  $0.28(5)$ & $0.313(2) $ & $0.318(8)$ & $0.311(2)$ 
& $0.311(3)$  & $0.309(4)$ \\ \hline 
\end{tabular}
\end{center}
\vskip -5mm
\caption{ Lattice spacing $a$ in units of the string tension $\sigma $ 
  for the values $\beta $; also shown $\alpha (\beta )$; $12^4$ 
lattice. }
\label{tab:1}
\vskip 5mm
\end{table*} 
We have calculated the potential with $\beta $ in the range $[9.4,10]$
on a $12^4$ lattice volume. The ``raw data'' are shown in
figure~\ref{fig:5} (left panel). From a fit of the raw data to eq.\
(\ref{eq:p2}), we have extracted the string tension in lattice units,
$\sigma a^2$, at each $\beta$ value (see table~\ref{tab:1}).  
In order to avoid any obfuscation by rotational symmetry breaking 
effects, only data from Wilson loops with $(100)$-orientation have been 
included.  
The obtained values for the 
string tensions can be used to express the static potential in
physical units, by first subtracting the self-energy $V_0(\beta )$
from the ``raw" potential in lattice units, and dividing the result by
$\sqrt{\sigma a^2}$.  The result is plotted for several $\beta$ values
as function of the distance in units of the physical string tension
$\sqrt{\sigma } \, r= n_s\, \sqrt{\sigma a^2}$ (see
figure~\ref{fig:5}, right panel). 
A fit of the data arising from 
all $\beta $-values to the potential $V(r)$ in (\ref{eq:p2}) is also 
included to guide the eye. 
We observe that the data at
different $\b$ values lies on the same line, and that rotational
symmetry breaking effects are small.

\subsection{Finite Temperature Transition}

    In SU(2) gauge-Higgs theory there is no unambiguous transition
between the Higgs phase and the temporary confinement phase, and
no local, gauge-invariant order parameter which can distinguish between 
the two phases (for a recent discussion, cf.\ \cite{ghiggs}).  In particular,
the center symmetry is trivial (i.e.\ $Z_{1}$), and Polyakov line VEVs
are non-zero throughout the phase diagram.  Nevertheless, there does exist
a finite temperature transition in this theory, where the Polyakov line
jumps from a small value to a larger value, and the linear potential at 
intermediate distances is lost.

    Denote the Polyakov line as
\beq
P(\vec{x}) \; = \; \prod _{t} U_4 (t, \vec{x}) \; . 
\label{eq:d2} 
\eeq
The standard order parameter for finite temperature transitions is
\beq
p(T) \; = \; \left\langle \frac{1}{N^3} \left\vert \sum _{\vec{x}} 
\tr \; P(\vec{x}) \right\vert \right\rangle 
\label{eq:d3} 
\eeq
on an $N^3 \times N_t$ lattice, where the temperature is 
$T = 1/(N_t \, a(\beta ))$. 
It was observed in refs.\ \cite{Pepe1,Pepe2} that G(2) lattice gauge theory 
undergoes a first order transition at a finite 
temperature, indicated by a jump of $p(T)$ at $T_c$.   As in SU(2)
gauge-Higgs theory, this first-order transition cannot be characterized
as a center symmetry-breaking transition.

      Consider a gauge transformation of the Polyakov line holonomy
\beq
P^{ \, \prime }  (\vec{x}) \; = \; \Omega (\vec{x},1) \, 
P (\vec{x}) \, \Omega ^T (\vec{x},1) \; , ~~~~~~ 
\Omega (\vec{x},1) \; = \;  \Omega (\vec{x},N_t) \; . 
\label{eq:d4} 
\eeq
By a judicious choice of gauge transformation $\Omega=T\Omega $,
where matrices $T$ and $\Omega$ are described in Appendix~\ref{sec:algebra}
and Appendix~\ref{sec:g2_sub}, respectively,
we may bring $P(x)$ into the form
\beq
T\; V \; P^{ \, \prime }  (\vec{x}) \; V^\dagger \; T^T \; = \; 
\left( \begin{array}{ccc} 
1 & 0 & 0 \\ 
0 & U_{ik} & 0 \\ 
0 & 0 & U^\ast _{lm} \end{array} \right) \; , ~~~~ 
U \; \in \; \mathrm{SU(3)} \; , 
\label{eq:d5} 
\eeq
which emphasizes the SU(3) subgroup.  The SU(3) subgroup is
of interest here because the trace of a Polyakov line is the sum of
its eigenvalues, the eigenvalues are elements of the Cartan subgroup,
and the Cartan subgroup of G(2) is the same as that of the SU(3) subgroup.
We have
\beq
\tr \;  P (\vec{x}) \; = \; 1 \; + \; 2 \, \mbox{Re} \; \tr U \; . 
\label{eq:dd8} 
\eeq
It is therefore sufficient to consider the trace of $3\times 3$ submatrix $U$.
If the Polyakov line VEV were exactly zero (the case of true confinement, rather
than temporary confinement), we would have
\beq
\Bigl\langle \frac{1}{3}  \; \mbox{Re} \; \tr U \Bigr\rangle 
\; = \; - \frac{1}{6} 
\label{eq:dd9} 
\eeq
exactly.  However, because of color screening, this relation can only be approximately true 
at low temperatures.  
\begin{figure*}[t]
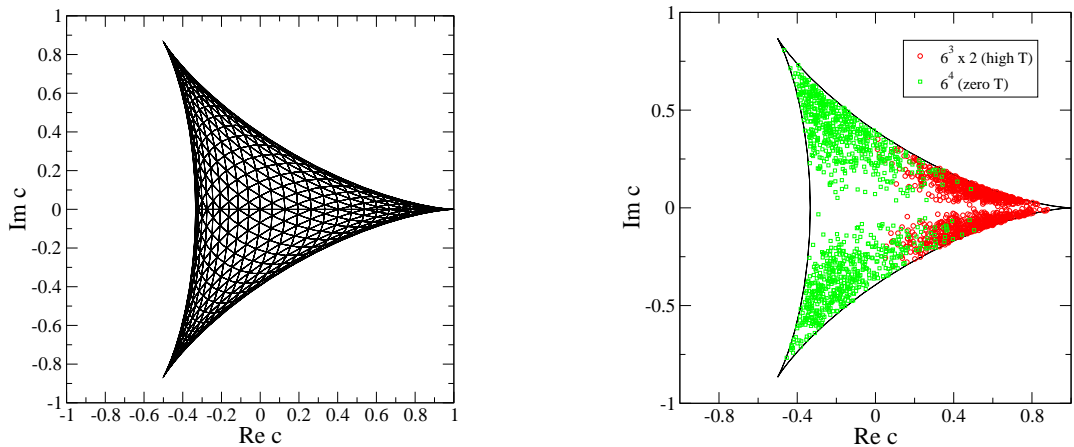

\epsfxsize=6cm 
\epsffile{pol_su3.eps} \hspace{2.0cm}
\epsfxsize=6cm 
\epsffile{pol_scatter6.eps} 
\caption{ Region of support for values $c (\vec{x})$ in (\ref{eq:d6}) 
for $\beta =0 $ (left panel). Right panel: Distribution of $c (\vec{x})$ 
at zero temperature (squares) and in the deconfinement phase 
(circles). 
}
\label{fig:d1}
\end{figure*}

   We define
\beq
c  (\vec{x})  =  \frac{1}{3} \tr  U  (\vec{x})  
\label{eq:d6} 
\eeq
which takes values within the curved triangle shown in 
figure~\ref{fig:d1}. The right panel of figure~\ref{fig:d1} shows 
the distribution of $c  (\vec{x})$ for $\beta=9.7$ at low and high
temperatures. At $\beta = 9.7$ and a $6^4$ lattice (low 
temperature case), we clearly observe that the points accumulate 
near the non-trivial center elements of SU(3), i.e., 
$$ 
c_+ \; = \; - \, 0.5 \; + \; i \,  \frac{\sqrt{3}}{2} \; , \hbo 
c_- \; = \; - \, 0.5 \; - \; i \,  \frac{\sqrt{3}}{2} \; . 
$$
As expected, the scatter plot is symmetric with respect to reflections 
across the $x$-axis. At temperatures $T>T_{c}$ ($6^{3}\times 2$ lattice
volume), a strong shift 
of the data away from the non-trivial center elements towards 
the unit element is clearly observed.  
    
\section{\label{sec:dominance}SU(3), $Z_{3}$, or SU(2) ``Dominance"?}

   Although the center of the G(2) gauge group is trivial, one may ask if, in some gauge, the infrared
dynamics is controlled by a subgroup of G(2) which \emph{does} have a center.  The motivation for this idea
goes back to 't Hooft's proposal for abelian projection \cite{thooft-mon}:  
In an SU(N) gauge theory, imagine imposing a gauge 
choice which leaves a remaining $U(1)^{N-1}$ gauge symmetry.  The resulting theory can be thought of as 
a gauge theory with a compact abelian gauge group, whose degrees of freedom are ``photons" and
monopoles, coupled to a set of ``matter" fields which consist of the gluons which are charged
with respect to the $U(1)^{N-1}$ gauge group.  If for some reason these gluons are very massive, then
the idea is that they are irrelevant to the infrared physics, which is controlled by an effective abelian
gauge theory.  This is known as ``abelian dominance".  Many numerical investigations have claimed to 
verify the existence of a large mass for the charged gluons.  Confinement is then supposed to be due
to the condensation of monopoles in the abelian theory.
 
    In the same spirit, let us consider fixing the gauge of the G(2) theory, in such a way that the action
remains invariant under an SU(3) subgroup of the G(2) gauge transformations.  The gauge-fixed theory
can be thought of as a theory of SU(3) gluons, coupled to vector matter fields (the remaining gluons of
G(2)) and ghosts.  If, for some unknown dynamical reason,  these vector matter and ghost fields aquire
a large effective mass, then the linear potential would be largely determined by pure SU(3) dynamics,
and $Z_{3}$ vortices associated with the center of the SU(3) subgroup would be expected to carry the
associated magnetic disorder.

    To investigate this idea, we turn to lattice Monte Carlo simulations.
We have found it useful, for the investigation of SU(3) dominance, to use the 
representation of the G(2) group by $7\times 7$ complex matrices, as presented in refs.\
\cite{Pepe2,Macfarlane}.  In this representation, any group element $\G \in G(2)$ can be expressed as
\beq
             \G = {\cal Z} {\cal U}
\eeq 
where ${\cal Z}$ is a $7\times 7$ unitary matrix which is a function of
a complex 3-vector $K$, and ${\cal U}$ is the matrix
\beq
            {\cal U} = \left( \begin{array}{ccc}
                          U &   & \cr
                            & 1 &  \cr
                            &   &  U^*  \end{array} \right)
\label{U}
\eeq
where $U$ is a $3\times 3$ SU(3) matrix.  The construction of the $\Z$ matrix
from a complex 3-vector is described, following ref.\ \cite{Pepe2}, in Appendix
\ref{sec:complex}.
The conjecture is that in a suitably chosen  ``maximal SU(3) gauge", which 
brings the $\Z$ matrices in the $\G=\Z\U$ link decomposition
as close as possible, on average, to the unit matrix, the information about confinement
in the intermediate distance regime is carried entirely by the $\U$
matrices.  This means that the string tension of Wilson
loops formed, in this gauge, from the $3\times 3$ $U$ matrices composing $\U$ would
approximately reproduce the full intermediate-distance string tension, while the
string tension of Wilson loops formed from the $\Z$ variables alone would vanish.
If SU(3) dominance of this kind
is exhibited by G(2) lattice gauge theory, then we can make a further reasonable conjecture,
which is that the confinement information in the $U$ matrices is carried
entirely by $Z_3$ center vortices.  If so, then (i) the $Z_3$ vortices by themselves should
also give a good estimate of the full intermediate string tension; and (ii) removing vortices from
the $\U$ matrices should completely eliminate the intermediate string tension in G(2) loop holonomies.
 
     We define maximal SU(3) gauge to be the gauge which minimizes
\bea
       R_1 &=& \sum_x \sum_{\m=1}^4 \left\{ \sum_{i=1}^3 \sum_{j=4}^7
          |u_\mu^{ij}(x)|^2 + \sum_{i=5}^7 \sum_{j=1}^4 |u_\mu^{ij}(x)|^2 \right.
\non \\
    & & \qquad  + \left. \left(\sum_{j=1}^3 + \sum_{j=5}^7 \right)|u_\mu^{4j}(x)|^2 \right\}
\eea
where $u_\m^{ij}(x)$ is the $(i,j)$ component of the G(2) link variable (note that $R_1=0$ for
$\Z=\Id_7$). It is straightforward
to check that multiplying a link variable on either the left or the right by 
an SU(3) subgroup element $\U$ leaves $R_1$ unchanged; it is therefore sufficient to use
only the $g=\Z$ transformations in fixing the gauge.  This also means that maximizing
$R_1$ leaves a residual local SU(3) symmetry, so that ``maximal SU(3)
gauge" is a good name for this gauge choice.
The next step is to use the remaining SU(3) gauge freedom to maximize
\bea
R_2 &=& \sum_{x} \sum_{\m=1}^4 \left\{
   \Bigl|u_\m^{11}(x) + u_\m^{22}(x) + u_\m^{33}(x)\Bigr|^2 \right.
\non \\
   & & \qquad + \left. \Bigl|u_\m^{55}(x) + u_\m^{66}(x) + u_\m^{77}(x)\Bigr|^2 \right\}
\eea
leaving a remnant local $Z_{3}$ symmetry in ``maximal $Z_3$ gauge."
This two-step process is reminiscent of fixing to indirect maximal
center gauge in SU(N) gauge theory.  Details regarding our Monte Carlo
updating procedure in the complex representation of G(2), and the method
of fixing to maximal SU(3) and maximal $Z_{3}$ gauges, 
can be found in Appendix \ref{sec:complex}.

\subsection{SU(3) and $\mathbf{Z_3}$ Projection}
 
   Having fixed to maximal $Z_3$ gauge, we compute the following observables:
\begin{enumerate}
\item {\bf Full Loops.}  Here we use the full G(2) link variables
$\G=\Z \U$ to compute Wilson loops, 
and of course gauge-fixing is irrelevant to the loop values.
\item {\bf SU(3) Projected Loops.}  These are loops computed from
SU(3) link variables $U$, extracted from the top $3\times 3$  block of 
the $7\times 7$ $\U$ link matrices
\beq
            {\cal U} = \left( \begin{array}{ccc}
                          U &   & \cr
                            & 1 &  \cr
                            &   &  U^*  \end{array} \right)
\eeq
\item {\bf $\mathbf{Z_3}$ Projected Loops.}   These loops are
just constructed from the $z_\m(x)\in Z_3$ link variables obtained 
from center-projection of the $3\times 3$ $U$-link variables. 
\item {\bf $\boldmath \Z$ Loops.} These ``SU(3)-suppressed" loops are
computed from the $\Z$ link variables alone; i.e.\ we set the SU(3) part of the $\G=\Z
\U$ link variables to $\U=\Id_7$.
\item {\bf G(2) Vortex-Only Loops.}  In the decomposition of G(2) link
variables $\G=\Z \U$, replace the $SU(3)$ factor $\U$ by the
vortex-only element 
\beq {\cal U} \ra V \equiv \left(
\begin{array}{ccc} zI_3 & & \cr & 1 & \cr & & z^*I_3 \end{array}
\right) 
\label{V}
\eeq 
and compute loops with the $7\times 7$ link variables
$\G=\Z V$.
\item {\bf G(2) Vortex-Removed Loops.}  Here we remove
vortices from the SU(3) elements $\U$ by the usual procedure \cite{dFE}, and use the modified,
vortex-removed $\U'$, together with $\Z$, to construct vortex-removed
G(2) lattice link variables $\G'=\Z \U'$.  In practice this
accomplished simply by multiplying the original $\G$ by $V^\dg$, i.e.\
construct loops from link variables
\beq 
\G' = \G V^\dg 
\eeq
\item {\bf SU(3) Vortex-Removed Loops.}  We construct loops from the SU(3)-projected,
vortex-removed lattice $U'=z^{*}U$.
\end{enumerate}

\begin{figure}[h!]
\includegraphics[width=8truecm]{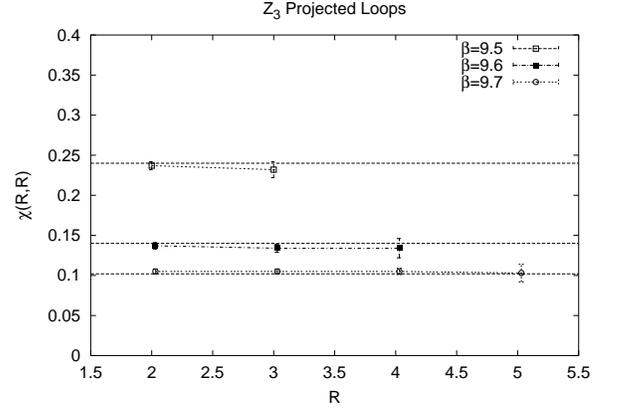}
\caption{Creutz ratios for $Z_3$-projected loops at $\b=9.5,9.6,9.7$. 
The horizontal lines show the values of the corresponding
intermediate string tensions, computed in section \ref{sec:static} on the unprojected lattice.}
\label{cproj}
\end{figure}

   Figure \ref{cproj} shows our results for Creutz
ratios of $Z_3$-projected configurations at couplings
$\b=9.5,~9.6,~9.7$. The straight lines shown are for the intermediate
string tension at these couplings, extracted from the unprojected lattice and listed
in Table \ref{tab:1} above.  The agreement of the
$Z_{3}$-projected string tension with the full intermediate string
tension is quite good.   

\begin{figure}[h!]
\includegraphics[width=8truecm]{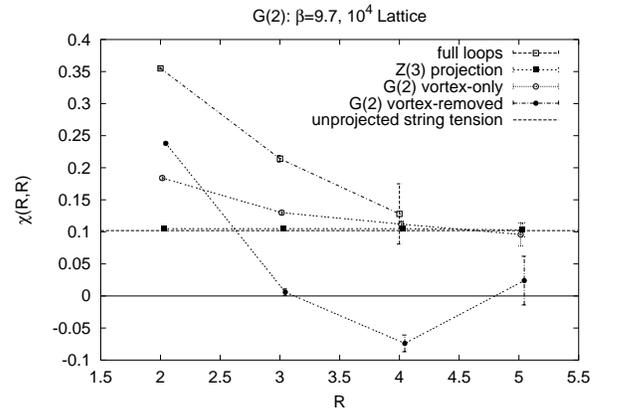}
\caption{Creutz ratios for the full, $Z_3$-projected, G(2) vortex-only,
and G(2) vortex-removed loops at $\b=9.7$}
\label{b97}
\end{figure}

   Figure \ref{b97} shows the data, at $\b=9.7$ on a $10^4$ lattice,
for Creutz ratios obtained from full, G(2) vortex-only, G(2)
vortex-removed, and $Z_{3}$ projected configurations.  Again, 
denoting the decomposition of the full links in maximal $Z_{3}$
gauge by $\G = \Z \U$, and the Z(3)-projection of 
$\U$ (eq.\ \ref{V}) by $V$,
then the G(2) vortex-only link variables are $\G_{vo} = \Z V$, and the
G(2) vortex-removed links are $\G_{vrem} = \Z \U V^\dg$.  
It is clear from the plot that the G(2) vortex-only
data appoaches the data obtained from full Wilson loops, while the 
linear potential is absent in the vortex-removed data.   

\begin{figure}[h!]
\includegraphics[width=8truecm]{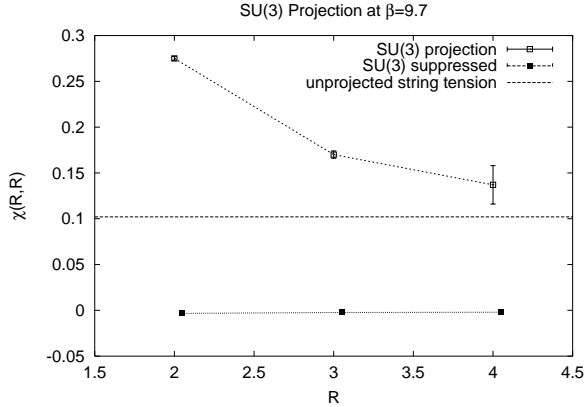}
\caption{Creutz ratios for SU(3) projected lattices, and for
G(2) lattices with the SU(3) link factors set to 
unity (``SU(3) suppressed").}
\label{su3}
\end{figure}

   If the confining disorder resides in the $Z_3$ degrees of freedom,
and these in turn lie in the SU(3) projected link variables, then 
SU(3) Dominance is implied. The
string tension on the unprojected lattice should be seen in the 
SU(3) projection, and removing SU(3) fluctuations from the G(2) link 
variables should send the
intermediate string tension to zero.  Both of these phenomena can be
seen in Fig.\ \ref{su3}. The suppression of the full string tension in
the $\Z$ loops, which
follows from suppression of SU(3) fluctuations, is particularly striking. 

\begin{figure}[h!]
\includegraphics[width=8truecm]{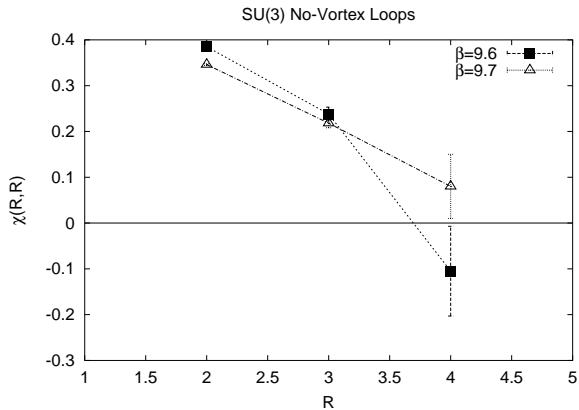}
\caption{Creutz ratios for SU(3)-projected, vortex-removed
Wilson loops, at $\b=9.6,~9.7$.}
\label{zs3}
\end{figure}

   Finally, we can remove vortices from the 
SU(3) link variables, and compute Creutz ratios from those lattices.
The results for $\b=9.6$ and $\b=9.7$ are displayed in Fig.\ \ref{zs3},
and appear to be consistent with a vanishing string tension for large loops.

\subsection{Problems with the Projections}
   
The success of SU(3) and $Z_{3}$-projected lattices in reproducing the
asymptotic string tension of G(2) gauge theory, together with the
vanishing string tension in ``vortex-removed" lattices, tends to
obscure the fact that these are, after all, gauge-dependent
observables.  That fact doesn't necessarily mean that, e.g., the
$Z_{3}$ vortices located via projection are unphysical, but neither do
the projected results alone prove that these are physical objects.
The really crucial numerical evidence in favor of the physical nature
of vortices, in SU(2) and SU(3) gauge theories, was the correlations
that were found found between P-vortex location and gauge-invariant
observables.  In particular

\begin{enumerate} 
\item Plaquettes at the location of P-vortices have, on the
  unprojected lattice, a plaquette action which is significantly
  higher than average.
\item A vortex-limited Wilson loop $W_{n}(C)$ is an unprojected Wilson
  loop, evaluated in an ensemble of configurations in which $n$
  P-vortices pierce the loop of the projected lattice.  It has been
  found that as the loop area becomes large 
  \beq \frac{W_{n}(C) }{
    W_{0}(C)} \ra e^{2\pi i n/N} 
  \eeq 
  where $N$ is the number of
  colors.  This is the expected behavior, if a P-vortex on the
  projected lattice locates a thick vortex on the unprojected lattice.
\item In an SU(N) gauge theory, a Wilson loop calculated in the
  ``vortex-removed" lattice can be equally well described as the
  expectation value of the product
\beq
          W_{vrem}(C) = \langle Z^{*}(C) \Tr[U(C)] \rangle
\eeq
where $Z(C),~U(C)$ are loop holonomies on the projected and
unprojected lattices, respectively.  If $W_{vrem}$ has a vanishing
asymptotic string tension, then this fact translates directly into a
correlation of the phase of the projected loop $Z(C)$, induced by
P-vortices, with the phase of the gauge-invariant observable
$\Tr[U(C)]$ on the unprojected lattice.
\end{enumerate}

Unlike the SU(2) and SU(3) cases, we have so far found no discernable
correlation between gauge-invariant G(2) Wilson loops, and the SU(3)
and $Z_{3}$ projected loops.  Plaquettes pierced by $Z_{3}$ vortices
have no higher action, on the unprojected lattice, than other
plaquettes, and the VEVs of vortex-limited Wilson loops $W_{n}(C)$
appear to be completely independent of the number of vortices $n$.
Although the string tension of vortex-removed loops vanishes in G(2)
gauge theory, the vortex-removed loops $W_{vrem}(C)$ cannot be
expressed as a product of vortex loops and gauge-invariant loops,
hence the VEV of $W_{vrem}(C)$ does not directly correlate the vortex
observables with gauge-invariant observables.  It should be noted that
in SU(N) gauge theories, vortex removal is a very minimal disturbance
of the lattice, especially at weak couplings.  Vortex removal affects
the plaquette action only at the location of P-plaquettes, whose
density falls exponentially with coupling $\b$.  In G(2) gauge theory,
matters are different.  In that case $Z_{3}$ vortex removal is a
serious butchery of the original configuration, affecting every
plaquette of the lattice.\footnote{In G(2) gauge theory, vortex removal 
is a matter of multiplying each link by a matrix diag$[z^{*}I_{3},1,zI_{3}]$ 
which is not a center element, and which does not commute with the
factor we have denoted by $\Z$.  In general this procedure will change the
plaquette action, in an uncontrolled way, at every plaquette on the lattice.
This is in complete contrast to SU(N) gauge theories, where multiplication of
link variables by projected center elements looks locally, almost everywhere,
like a $Z_{N}$ gauge transformation, and changes plaquette actions only at the 
location of P-plaquettes.} 

   For these reasons, we think that the results of SU(3) and $Z_{3}$
projection in G(2) gauge theory are misleading.  In particular, the
supposed vortices located by $Z_{3}$ projection appear to not affect
any local gauge-invariant observables (i.e. Wilson loops), and on
these grounds we believe that the objects identified by $Z_{3}$
projection are simply unphysical.   For these reasons, we think that the results of SU(3) and $Z_{3}$
projection in G(2) gauge theory are misleading.  In particular, the
supposed vortices located by $Z_{3}$ projection appear to not affect
any local gauge-invariant observables (i.e. Wilson loops), and on
these grounds we believe that the objects identified by $Z_{3}$
projection are simply unphysical.\footnote{A possibly related issue 
has been studied recently by Pepe and Wiese \cite{PW}.  
These authors simulate G(2) lattice gauge-Higgs theory, where the Higgs field can
spontaneously break the symmetry down to SU(3).  An expectation value for the
Higgs field gives a mass to six of the fourteen G(2) gluons, while eight other gluons, associated
with the SU(3) subgroup, remain massless.  The masses of the massive gluons  can be
adjusted, from zero to infinity, by the hopping
parameter in the Higgs Lagrangian, and in this way the gauge-Higgs theory can interpolate 
between pure G(2) and pure SU(3) gauge theory.  The question addressed in ref.\ \cite{PW} is 
whether the deconfinement transition of
SU(3) gauge theory, which is certainly a $Z_{3}$ symmetry-breaking transition, is
continuously connected, in the phase diagram, to the first-order finite-temperature transition in
G(2) gauge theory.  The answer appears to be no, and the authors argue that in fact the
two transitions have different origins.  This conclusion seems consistent with our finding
that observables defined in SU(3) and $Z_{3}$ projections  appear to be physically irrelevant in 
pure G(2) gauge theory.}

\FIGURE[t!]{
\includegraphics[width=7truecm]{pot_su2.eps}
\caption{The G(2) static potential, compared to the static potentials
computed in SU(2)-only and SU(2)-removed lattices, at $\b=9.6$.  String tensions
$s=\s a^{2}$ are extracted from fits to eq.\ \rf{eq:p2}.} 
\label{pot-su2}
}

One might ask if the reason for the ultimate failure of the SU(3) and
$Z_{3}$ projections lies in the choice of the subgroup.  If we would
calculate Wilson loops from the $7\times 7$ SU(3)-projected link
variables $\U$, rather than the $3\times 3$ SU(3) link variables $U$,
then those Wilson loops would not have an area law, due to the unit
element in the $7\times 7$ $\U$ matrix; i.e.\ the existence of an
element of the fundamental representation of G(2) which transforms as
a singlet under the SU(3) subgroup.  Suppose that we instead look for
the smallest subgroup of G(2) such that all elements of the
fundamental representation of G(2) transform as non-singlets under
that subgroup.  It turns out that the SU(2) subgroup defined by the
generators $C_{8},C_{9},C_{10}$ (see Appendix \ref{sec:algebra})
satisfies this condition; the fundamental representation of G(2)
transforms, under this subgroup, as a triplet and two doublets. 

We then impose a ``maximal SU(2)" gauge, and calculate the Wilson
loops of the SU(2)-only and SU(2)-removed link variables.  The results
are impressive, and are displayed in Fig.\ \ref{pot-su2}.  However, up to now
we have not found any significant correlation between 
SU(2)-only and G(2) Wilson loops, and in this respect the SU(2) projection appears
to be also problematic.
 
We believe that the reason that SU(3), $Z_{3}$, and SU(2)-projected
loops apparently fail to correlate with gauge-invariant observables
is that these projections have no gauge-invariant motivation from the
beginning.  We may reasonably suppose that a gauge theory with gauge
group $G$ is ``dominated" by the degrees of freedom associated with a
subgroup $H$, if, independent of any gauge-fixing, the gauge-invariant
string tension depends only on the representation of the subgroup.
This is the case for the $Z_{N}$ center subgroup of SU(N) gauge
theories, at large distance scales.  Center projection is then a
method for isolating the large-scale confining fluctuations.  In
contrast, the SU(3), $Z_{3}$, and SU(2) subgroups of G(2) have no
relation to G(2) Casimir scaling, at intermediate distances, nor are
any of these subgroups singled out at large scales, where the string
tension vanishes for any representation.\footnote{The situation would
be different in a G(2) gauge-Higgs theory, with G(2) broken to
SU(3).}  In fact, similar remarks can be made regarding abelian
projection in pure SU(N) gauge theories, which we used to motivate the
subgroup projections of this section.  The abelian U(1)${}^{N-1}$
subgroup cannot easily account for Casimir scaling at intermediate
distances, while at asymptotic distances the string tension of an
abelian-projected loop in fact depends only on the N-ality, rather than the
abelian charge of the loop \cite{j3}.

\section{\label{sec:conclusions}Conclusions}

In this article we have presented a unified picture of vacuum
fluctuations in G(2) and SU(N) gauge theories, in which a
two-dimensional surface slice of the four-dimensional lattice can be
decomposed into domains corresponding to elements of the gauge group
center.  Domains associated with center elements different from the
identity are just two-dimensional cross-sections of the usual $Z_{N}$
vortices; the new feature is that we also allow for domains associated
with the unit center element, which exists in G(2) as well as SU(N).
This domain structure accounts for the representation dependence of
the asymptotic string tension on the center subgroup, which in the
case of G(2) implies a vanishing asymptotic string tension.  The
existence and Casimir scaling of the linear potential at intermediate
distances is explained by random spatial variations of the color
magnetic flux in the interior of the center domains.

The picture we have outlined is a kind of phenomenology.  Domain
structure is arrived at by simply asking what features of typical
vacuum fluctuations, in a Euclidean 4-volume, can account for both
Casimir scaling of intermediate string tensions, and color-screening
in the asymptotic string tensions.  Of course one would like to
support this picture by either analytical methods or numerical
simulations.  There is, in fact, abundant numerical evidence for the
relevance of $Z_{2}$ and $Z_{3}$ center vortices in SU(2) and SU(3)
lattice gauge theories \cite{review,michael}, but unfortunately the
usual tools which are used to locate such vortices in lattice
simulations (i.e.\ center projection in maximal center gauge) are not
of much use in locating domains associated with the unit element.  At
the moment we can only appeal to a number of analytical studies of the
one-loop effective action of vortex configurations, which show that
this action is stationary for magnetic flux quantized in units
corresponding to the gauge group center elements, including the unit
element \cite{Hugo,Diakonov,Bordag}.  Thus the dynamics underlying
center vortex formation may ultimately be the same for the unit and
non-unit element varieties.  The picture we have presented for Casimir
scaling at intermediate distance scales is motivated by the notion of
dimensional reduction, and by various approximate treatments of the
Yang-Mills vacuum wavefunctional \cite{Me1,Me2,Mansfield,Nair,Leigh,Freidel}.

We have also calculated certain static properties of G(2) lattice
gauge theory, namely, the intermediate string tension, and the
distribution of Polyakov line holonomies at zero and finite
temperature.  We see very clear evidence of a linear potential in the
fundamental representation, but we have not attempted to compute the
potential for higher representations, needed to verify Casimir
scaling, or the operator-mixings needed to see string breaking.  These
are left for future investigations.  The eigenvalues of the Polyakov
line are those of the SU(3) subgroup, and we have shown that the
finite temperature phase transition is accompanied by a shift in the
concentration of the Polyakov line eigenvalues from non-trivial to
trivial SU(3) center elements.

Finally, we have investigated projection of G(2) link variables to
various subgroups.  These projections, to SU(3), $Z_{3}$, and SU(2)
subgroups, so far lack the gauge-invariant motivation that 
exists for center projection in SU(N) gauge theories, where the representation
dependence of the asymptotic string tension is given entirely by the
representation of the $Z_{N}$ subgroup.  The SU(3), $Z_{3}$, and SU(2)
subgroups of G(2) have no natural connection to either G(2) Casimir
scaling at intermediate distances, where a non-zero string tension exists, 
or to the asymptotic regime, where the string tension vanishes.  The fact that we 
have found no correlation whatever, between the projected Wilson loops and
gauge-invariant observables in G(2) gauge theory, probably reflects
the fact that these subgroups do not appear naturally in any
gauge-invariant set of observables.  The only subgroup which really
does appear to be relevant to pure non-abelian gauge theories is the
center subgroup, and this relevance persists even in cases such as
G(2) gauge theory, where the center subgroup consists of only a single
unit element.

%
\acknowledgments{%
We thank M.\ Pepe, M. Quandt, and U.-J.\ Wiese for helpful discussions. 
Our research is supported in part by the U.S. Department of Energy
under Grant No.\ DE-FG03-92ER40711 (J.G.), the Slovak Science
and Technology Assistance Agency under Contract 
No.\  APVT-51-005704 (\v{S}.O.), and the Deutsche Forschungsgemeinschaft
under contract DFG-Re856/4-2,3.}

\appendix
\section{\label{sec:algebra} G(2) algebra}

In the following, we will study the fundamental representation of G(2) 
in more detail. In this representation, the generators can be chosen as 
real  $7 \times 7$ matrices. The group elements $U$ satisfy the constraints 
\beq 
U \, U^T \; = \; 1 \; , \hbo 
T_{abc} \; = \; T_{def} \; U_{da} \; U_{eb} \; U_{fc} \; , 
\label{eq:constr1} 
\eeq
where $T_{abc}$ is a total anti-symmetric tensor~\cite{Pepe1}. 
The group elements are generated by the Lie algebra: 
\beq
U \; = \; \exp \left\{ -i \, \sum _{a=1}^{14} \rho _a \; i C_a \right\} \; , 
\hbo \rho \in \R^{14} \; . 
\label{eq:al1} 
\eeq 
An explicit realization of the generators $i \, C_a$ was presented 
in~\cite{Cacciatori:2005yb}. For the readers convenience, these 
matrices are provided below. 
For this choice of representation, these elements are 
\beq
T_{123} \; = \; T_{145} \; = \; T_{176} \; = \; 
T_{246} \; = \; T_{257} \; = \; T_{347}  \; = \; T_{365}  \; = \; 1 \; . 
\label{eq:constr2} 
\eeq 
All linear combinations of the generators $C_a$ form the vector space 
$L$ of the Lie algebra. 

\begin{widetext}
    The embedding of $G(2)$ in the group $SO(7)$ is 
generated by the $14$ Lie algebra elements $C_k, k = 1, \ldots , 14$: 
\bea
C_1 &=&\left(
\begin{array}{ccccccc}
0 & 0 & 0 & 0 & 0 & 0 & 0 \\
0 & 0 & 0 & 0 & 0 & 0 & 0 \\
0 & 0 & 0 & 0 & 0 & 0 & 0 \\
0 & 0 & 0 & 0 & 0 & 0 & -1 \\
0 & 0 & 0 & 0 & 0 & -1 & 0 \\
0 & 0 & 0 & 0 & 1 & 0 & 0 \\
0 & 0 & 0 & 1 & 0 & 0 & 0 
\end{array}
\right)
\qquad
 C_2 =\left(
\begin{array}{ccccccc}
0 & 0 & 0 & 0 & 0 & 0 & 0 \\
0 & 0 & 0 & 0 & 0 & 0 & 0 \\
0 & 0 & 0 & 0 & 0 & 0 & 0 \\
0 & 0 & 0 & 0 & 0 & 1 & 0 \\
0 & 0 & 0 & 0 & 0 & 0 & -1 \\
0 & 0 & 0 & -1 & 0 & 0 & 0 \\
0 & 0 & 0 & 0 & 1 & 0 & 0 
\end{array}
\right)
\non \\
C_3 &=&\left(
\begin{array}{ccccccc}
0 & 0 & 0 & 0 & 0 & 0 & 0 \\
0 & 0 & 0 & 0 & 0 & 0 & 0 \\
0 & 0 & 0 & 0 & 0 & 0 & 0 \\
0 & 0 & 0 & 0 & -1 & 0 & 0 \\
0 & 0 & 0 & 1 & 0 & 0 & 0 \\
0 & 0 & 0 & 0 & 0 & 0 & -1 \\
0 & 0 & 0 & 0 & 0 & 1 & 0 
\end{array}
\right)
\qquad
C_4 =\left(
\begin{array}{ccccccc}
0 & 0 & 0 & 0 & 0 & 0 & 0 \\
0 & 0 & 0 & 0 & 0 & 0 & 1 \\
0 & 0 & 0 & 0 & 0 & 1 & 0 \\
0 & 0 & 0 & 0 & 0 & 0 & 0 \\
0 & 0 & 0 & 0 & 0 & 0 & 0 \\
0 & 0 & -1 & 0 & 0 & 0 & 0 \\
0 & -1 & 0 & 0 & 0 & 0 & 0 
\end{array}
\right)
\non \\
C_5 &=&\left(
\begin{array}{ccccccc}
0 & 0 & 0 & 0 & 0 & 0 & 0 \\
0 & 0 & 0 & 0 & 0 & -1 & 0 \\
0 & 0 & 0 & 0 & 0 & 0 & 1 \\
0 & 0 & 0 & 0 & 0 & 0 & 0 \\
0 & 0 & 0 & 0 & 0 & 0 & 0 \\
0 & 1 & 0 & 0 & 0 & 0 & 0 \\
0 & 0 & -1 & 0 & 0 & 0 & 0 
\end{array}
\right)
\qquad
C_6 =\left(
\begin{array}{ccccccc}
0 & 0 & 0 & 0 & 0 & 0 & 0 \\
0 & 0 & 0 & 0 & 1 & 0 & 0 \\
0 & 0 & 0 & -1 & 0 & 0 & 0 \\
0 & 0 & 1 & 0 & 0 & 0 & 0 \\
0 & -1 & 0 & 0 & 0 & 0 & 0 \\
0 & 0 & 0 & 0 & 0 & 0 & 0 \\
0 & 0 & 0 & 0 & 0 & 0 & 0 
\end{array}
\right)
\non \\
C_7 &=&\left(
\begin{array}{ccccccc}
0 & 0 & 0 & 0 & 0 & 0 & 0 \\
0 & 0 & 0 & -1 & 0 & 0 & 0 \\
0 & 0 & 0 & 0 & -1 & 0 & 0 \\
0 & 1 & 0 & 0 & 0 & 0 & 0 \\
0 & 0 & 1 & 0 & 0 & 0 & 0 \\
0 & 0 & 0 & 0 & 0 & 0 & 0 \\
0 & 0 & 0 & 0 & 0 & 0 & 0 
\end{array}
\right)
\qquad
C_8 =\frac 1{\sqrt 3} \left(
\begin{array}{ccccccc}
0 & 0 & 0 & 0 & 0 & 0 & 0 \\
0 & 0 & -2 & 0 & 0 & 0 & 0 \\
0 & 2 & 0 & 0 & 0 & 0 & 0 \\
0 & 0 & 0 & 0 & 1 & 0 & 0 \\
0 & 0 & 0 & -1 & 0 & 0 & 0 \\
0 & 0 & 0 & 0 & 0 & 0 & -1 \\
0 & 0 & 0 & 0 & 0 & 1 & 0 
\end{array}
\right)
\non \\
C_9 &=&\frac 1{\sqrt 3} \left(
\begin{array}{ccccccc}
0 & -2 & 0 & 0 & 0 & 0 & 0 \\
2 & 0 & 0 & 0 & 0 & 0 & 0 \\
0 & 0 & 0 & 0 & 0 & 0 & 0 \\
0 & 0 & 0 & 0 & 0 & 0 & 1 \\
0 & 0 & 0 & 0 & 0 & -1 & 0 \\
0 & 0 & 0 & 0 & 1 & 0 & 0 \\
0 & 0 & 0 & -1 & 0 & 0 & 0 
\end{array}
\right)
\qquad
C_{10} =\frac 1{\sqrt 3} \left(
\begin{array}{ccccccc}
0 & 0 & -2 & 0 & 0 & 0 & 0 \\
0 & 0 & 0 & 0 & 0 & 0 & 0 \\
2 & 0 & 0 & 0 & 0 & 0 & 0 \\
0 & 0 & 0 & 0 & 0 & -1 & 0 \\
0 & 0 & 0 & 0 & 0 & 0 & -1 \\
0 & 0 & 0 & 1 & 0 & 0 & 0 \\
0 & 0 & 0 & 0 & 1 & 0 & 0 
\end{array}
\right)
\non \\
 C_{11} &=& \frac 1{\sqrt 3} \left(
\begin{array}{ccccccc}
0 & 0 & 0 & -2 & 0 & 0 & 0 \\
0 & 0 & 0 & 0 & 0 & 0 & -1 \\
0 & 0 & 0 & 0 & 0 & 1 & 0 \\
2 & 0 & 0 & 0 & 0 & 0 & 0 \\
0 & 0 & 0 & 0 & 0 & 0 & 0 \\
0 & 0 & -1 & 0 & 0 & 0 & 0 \\
0 & 1 & 0 & 0 & 0 & 0 & 0 
\end{array}
\right)
\qquad
C_{12} = \frac 1{\sqrt 3} \left(
\begin{array}{ccccccc}
0 & 0 & 0 & 0 & -2 & 0 & 0 \\
0 & 0 & 0 & 0 & 0 & 1 & 0 \\
0 & 0 & 0 & 0 & 0 & 0 & 1 \\
0 & 0 & 0 & 0 & 0 & 0 & 0 \\
2 & 0 & 0 & 0 & 0 & 0 & 0 \\
0 & -1 & 0 & 0 & 0 & 0 & 0 \\
0 & 0 & -1 & 0 & 0 & 0 & 0 
\end{array}
\right)
\non \\
C_{13} &=& \frac 1{\sqrt 3} \left(
\begin{array}{ccccccc}
0 & 0 & 0 & 0 & 0 & -2 & 0 \\
0 & 0 & 0 & 0 & -1 & 0 & 0 \\
0 & 0 & 0 & -1 & 0 & 0 & 0 \\
0 & 0 & 1 & 0 & 0 & 0 & 0 \\
0 & 1 & 0 & 0 & 0 & 0 & 0 \\
2 & 0 & 0 & 0 & 0 & 0 & 0 \\
0 & 0 & 0 & 0 & 0 & 0 & 0 
\end{array}
\right)
\qquad
C_{14} = \frac 1{\sqrt 3} \left(
\begin{array}{ccccccc}
0 & 0 & 0 & 0 & 0 & 0 & -2 \\
0 & 0 & 0 & 1 & 0 & 0 & 0 \\
0 & 0 & 0 & 0 & -1 & 0 & 0 \\
0 & -1 & 0 & 0 & 0 & 0 & 0 \\
0 & 0 & 1 & 0 & 0 & 0 & 0 \\
0 & 0 & 0 & 0 & 0 & 0 & 0 \\
2 & 0 & 0 & 0 & 0 & 0 & 0 
\end{array}
\right)
\eea
The two matrices $C_3$ and $C_8$ generate the Cartan subgroup of $G(2)$.
Furthermore, there are 6 $SU(2)$ subgroups generated by the elements:
\begin{eqnarray}
\begin{array}{llll}
C_1, & C_2, 
& Y_1 := C_3, 
& [C_1,C_2] = 2 Y_1, \\
C_4, & C_5, 
& Y_2 := \frac{1}{2} \left( \sqrt{3} C_8 + C_3 \right), 
& [C_4,C_5] = 2 Y_2 \\
C_6, & C_7, 
& Y_3 := \frac{1}{2} \left( - \sqrt{3} C_8 + C_3 \right), 
& [C_6,C_7] = 2 Y_3 \\
\sqrt{3} C_9, & \sqrt{3} C_{10}, 
& Y_4 := \sqrt{3} C_8, 
& [\sqrt{3} C_9, \sqrt{3} C_{10}] = 2 Y_4  \\
\sqrt{3} C_{11}, & \sqrt{3} C_{12}, 
& Y_5 := \frac{1}{2} \left( - \sqrt{3} C_8 + 3 C_3 \right), 
& [\sqrt{3} C_{11}, \sqrt{3} C_{12}] = 2 Y_5 \\
\sqrt{3} C_{13}, & \sqrt{3} C_{14}, 
& Y_6 := \frac{1}{2} \left( \sqrt{3} C_8 + 3 C_3 \right), 
& [\sqrt{3} C_{13}, \sqrt{3} C_{14}] = 2 Y_6
\label{algebra77}
\end{array} \, .
\end{eqnarray}
 The first three $SU(2)$ subgroups form $4$ dimensional real
representations of $SU(2)$ as subgroups of $SO(7)$ and they 
generate an $SU(3)$ subgroup of $G(2)$. The representations of the 
remaining three $SU(2)$ subgroups (as subgroups of $SO(7)$) are a 
little bit more complicated - they are $7$ dimensional but they are
reducible: the representation space is the direct sum of a $3$
dimensional adjoint representation and a $4$ dimensional real
representation (as in the case of the first three $SU(2)$ subgroups).

The various $SU(2)$ subgroups of $G(2) \subset SO(7)$ are obtained
by exponentiation: 
\bea
\lefteqn{\exp(\alpha_1 C_1 + \alpha_2 C_2 +\alpha_3 Y_1)}
\non \\ 
   &=& \left(
\begin{array}{ccccccc}
1 & 0 & 0 & 0 & 0 & 0 & 0 \\
0 & 1 & 0 & 0 & 0 & 0 & 0 \\
0 & 0 & 1 & 0 & 0 & 0 & 0 \\
0 & 0 & 0 & 
\cos \alpha                     & - \hat \alpha_3 \sin \alpha   & 
\hat \alpha_2 \sin \alpha       & - \hat \alpha_1 \sin \alpha   \\
0 & 0 & 0 & 
\hat \alpha_3 \sin \alpha       & \cos \alpha                   & 
- \hat \alpha_1 \sin \alpha     & - \hat \alpha_2 \sin \alpha   \\
0 & 0 & 0 & 
- \hat \alpha_2 \sin \alpha     & \hat \alpha_1 \sin \alpha     & 
\cos \alpha                     & - \hat \alpha_3 \sin \alpha   \\
0 & 0 & 0 & 
\hat \alpha_1 \sin \alpha       & \hat \alpha_2 \sin \alpha     & 
\hat \alpha_3 \sin \alpha       & \cos \alpha  
\end{array}
\right) = 
\left(
\begin{array}{ccccccc}
1 & 0 & 0 & 0 & 0 & 0 & 0 \\
0 & 1 & 0 & 0 & 0 & 0 & 0 \\
0 & 0 & 1 & 0 & 0 & 0 & 0 \\
0 & 0 & 0 & a_0 & -a_3 & a_2 & -a_1 \\
0 & 0 & 0 & a_3 & a_0 & -a_1 & -a_2 \\
0 & 0 & 0 & -a_2 & a_1 & a_0 & -a_3 \\
0 & 0 & 0 & a_1 & a_2 & a_3 & a_0 
\end{array}
\right)
\non \\ \non \\
\lefteqn{\exp(\alpha_1 C_4 + \alpha_2 C_5 +\alpha_3 Y_2)} 
\non \\ 
  &=& \left(
\begin{array}{ccccccc}
1 & 0 & 0 & 0 & 0 & 0 & 0 \\
0 & 
\cos \alpha                     & - \hat \alpha_3 \sin \alpha   &
0 & 0 & 
- \hat \alpha_2 \sin \alpha     & \hat \alpha_1 \sin \alpha     \\
0 & 
\hat \alpha_3 \sin \alpha       & \cos \alpha                   &
0 & 0 &
\hat \alpha_1 \sin \alpha       & \hat \alpha_2 \sin \alpha     \\
0 & 0 & 0 & 1 & 0 & 0 & 0 \\
0 & 0 & 0 & 0 & 1 & 0 & 0 \\
0 & 
\hat \alpha_2 \sin \alpha       & - \hat \alpha_1 \sin \alpha   & 
0 & 0 &
\cos \alpha                     & - \hat \alpha_3 \sin \alpha   \\
0 &
- \hat \alpha_1 \sin \alpha     & - \hat \alpha_2 \sin \alpha   & 
0 & 0 & 
\hat \alpha_3 \sin \alpha       & \cos \alpha  
\end{array}
\right) = 
\left(
\begin{array}{ccccccc}
1       & 0     & 0     & 0     & 0     & 0     & 0     \\
0       & a_0   & -a_3  & 0     & 0     & -a_2  & a_1   \\
0       & a_3   & a_0   & 0     & 0     & a_1   & a_2   \\
0       & 0     & 0     & 1     & 0     & 0     & 0     \\
0       & 0     & 0     & 0     & 1     & 0     & 0     \\
0       & a_2   & -a_1  & 0     & 0     & a_0   & -a_3  \\
0       & -a_1  & -a_2  & 0     & 0     & a_3   & a_0   \\
\end{array}
\right) 
\non \\ \non \\
\lefteqn{\exp(\alpha_1 C_6 + \alpha_2 C_7 +\alpha_3 Y_3) =} 
\non \\
  &=& \left(
\begin{array}{ccccccc}
1 & 0 & 0 & 0 & 0 & 0 & 0 \\
0 & 
\cos \alpha                     & \hat \alpha_3 \sin \alpha     &
- \hat \alpha_2 \sin \alpha     & \hat \alpha_1 \sin \alpha     &
0 & 0 \\
0 & 
- \hat \alpha_3 \sin \alpha     & \cos \alpha                   &
- \hat \alpha_1 \sin \alpha     & - \hat \alpha_2 \sin \alpha   &
0 & 0 \\
0 & 
\hat \alpha_2 \sin \alpha       & \hat \alpha_1 \sin \alpha     & 
\cos \alpha                     & - \hat \alpha_3 \sin \alpha   &
0 & 0 \\
0 &
- \hat \alpha_1 \sin \alpha     & \hat \alpha_2 \sin \alpha     & 
\hat \alpha_3 \sin \alpha       & \cos \alpha                   &
0 & 0 \\
0 & 0 & 0 & 0 & 0 & 1 & 0 \\
0 & 0 & 0 & 0 & 0 & 0 & 1 \\
\end{array}
\right) = 
\left(
\begin{array}{ccccccc}
1       & 0     & 0     & 0     & 0     & 0     & 0     \\
0       & a_0   & a_3   & -a_2  & a_1   & 0     & 0     \\
0       & -a_3  & a_0   & -a_1  & -a_2  & 0     & 0     \\
0       & a_2   & a_1   & a_0   & -a_3  & 0     & 0     \\
0       & -a_1  & a_2   & a_3   & a_0   & 0     & 0     \\
0       & 0     & 0     & 0     & 0     & 1     & 0     \\
0       & 0     & 0     & 0     & 0     & 0     & 1     \\
\end{array}
\right) 
\non \\ \non \\
\lefteqn{\exp(\alpha_1 \sqrt{3} C_9 + \alpha_2 \sqrt{3} C_{10} +\alpha_3 Y_4)} 
\non \\ 
  &=& \left(
\begin{array}{cc} 
A & 0 \\ 0 & B 
\end{array}
\right)  
\eea
where
\beq
\alpha = \sqrt{\alpha_1^2+\alpha_2^2+\alpha_3^2} ~~~,~~~ 
\hat \alpha_k = \frac{\alpha_k }{ \alpha}~~~,~~~ a_0 = \cos \alpha ~~~,~~~ 
a_k = \hat \alpha_k \sin \alpha
\eeq 
and
\bea
A  &=&  \left(
\begin{array}{ccc}
2 \cos^2 \alpha - 1 + 2 \hat \alpha_3 \hat \alpha_3 \sin^2 \alpha & 
- \hat \alpha_1 \sin 2 \alpha - \hat \alpha_2 \hat \alpha_3 \sin^2 \alpha & 
- \hat \alpha_2 \sin 2 \alpha + \hat \alpha_1 \hat \alpha_3 \sin^2 \alpha 
\\
\hat \alpha_1 \sin 2 \alpha - \hat \alpha_2 \hat \alpha_3 \sin^2 \alpha& 
2 \cos^2 \alpha - 1 + 2 \hat \alpha_2 \hat \alpha_2 \sin^2 \alpha &  
- \hat \alpha_3 \sin 2 \alpha - \hat \alpha_1 \hat \alpha_2 \sin^2 \alpha 
\\
\hat \alpha_2 \sin 2 \alpha + \hat \alpha_1 \hat \alpha_3 \sin^2 \alpha & 
\hat \alpha_3 \sin 2 \alpha - \hat \alpha_1 \hat \alpha_2 \sin^2 \alpha & 
2 \cos^2 \alpha - 1 + 2 \hat \alpha_1 \hat \alpha_1 \sin^2 \alpha 
\end{array}
\right)
\non \\
B &=& \left(
\begin{array}{cccc}
\cos \alpha                     & \hat \alpha_3 \sin \alpha     &
- \hat \alpha_2 \sin \alpha     & \hat \alpha_1 \sin \alpha     \\
- \hat \alpha_3 \sin \alpha     & \cos \alpha                   &
- \hat \alpha_1 \sin \alpha     & - \hat \alpha_2 \sin \alpha   \\
\hat \alpha_2 \sin \alpha       & \hat \alpha_1 \sin \alpha     & 
\cos \alpha                     & - \hat \alpha_3 \sin \alpha   \\
- \hat \alpha_1 \sin \alpha     & \hat \alpha_2 \sin \alpha     & 
\hat \alpha_3 \sin \alpha       & \cos \alpha               
\end{array}
\right) 
\eea
The elements of the two remaining $SU(2)$ subgroups look quite similar.

   Alternatively, one may equally well use a complex representation of $G(2)$, as in
refs.\ \cite{Pepe1,Pepe2} and in section \ref{sec:dominance} of this article. The two
representations can be related by a similarity transformation 
$\tilde C_k = V C_k V^\dagger$, where $V$ is the unitary matrix
\bea
V &=&
\frac{1}{\sqrt{2}} \left(
\begin{array}{ccccccc}
\sqrt{2}& 0     & 0     & 0     & 0     & 0     & 0     \\
0       & -i    & -1    & 0     & 0     & 0     & 0     \\
0       & -1    & -i    & 0     & 0     & 0     & 0     \\
0       & 0     & 0     & i     & 1     & 0     & 0     \\
0       & 0     & 0     & 1     & i     & 0     & 0     \\
0       & 0     & 0     & 0     & 0     & i     & -1    \\
0       & 0     & 0     & 0     & 0     & 1     & -i     
\label{eq:ad1} 
\end{array}
\right) 
=  \frac{1}{\sqrt{2}} \left(
\begin{array}{cccc}
\sqrt{2}        & 0     & 0     & 0     \\
0       & -i \Id - \sigma_1     & 0     & 0     \\
0       & 0     & i \Id + \sigma_1      & 0     \\
0       & 0     & 0     & i \sigma_3-i \sigma_2  \\
\end{array}
\right)
\end{eqnarray}
and in particular
\begin{eqnarray}
\tilde C_1 &=& 
\left(
\begin{array}{cccc}
0       & 0     & 0     & 0     \\
0       & 0     & 0     & 0     \\
0       & 0     & 0     & i \sigma_3     \\
0       & 0     & i \sigma_3     & 0 \\
\end{array}
\right) \, , \, 
\tilde C_2 = 
\left(
\begin{array}{cccc}
0       & 0     & 0     & 0     \\
0       & 0     & 0     & 0     \\
0       & 0     & 0     & \Id   \\
0       & 0     & -\Id   & 0    \\
\end{array}
\right) \, , \, 
\tilde C_3 = 
\left(
\begin{array}{cccc}
0       & 0     & 0     & 0     \\
0       & 0     & 0     & 0     \\
0       & 0     & - i \sigma_3     & 0  \\
0       & 0     & 0   & i \sigma_3      \\
\end{array}
\right) \, , \, 
\non \\
\tilde C_4 &=& 
\left(
\begin{array}{cccc}
0       & 0     & 0     & 0     \\
0       & 0     & 0     & i \sigma_3     \\
0       & 0     & 0     & 0     \\
0       & i \sigma_3    & 0     & 0 \\
\end{array}
\right) \, , \, 
\tilde C_5 = 
\left(
\begin{array}{cccc}
0       & 0     & 0     & 0     \\
0       & 0     & 0     & \Id   \\
0       & 0     & 0     & 0     \\
0       & -\Id  & 0     & 0     \\
\end{array}
\right) \, , \, 
\tilde C_6 = 
\left(
\begin{array}{cccc}
0       & 0     & 0     & 0     \\
0       & 0     & i \sigma_3    & 0     \\
0       & i \sigma_3    & 0     & 0     \\
0       & 0     & 0     & 0     \\
\end{array}
\right) \, , \, 
\non \\
\tilde C_7 &=& 
\left(
\begin{array}{cccc}
0       & 0     & 0     & 0     \\
0       & 0     & -\Id  & 0     \\
0       & \Id   & 0     & 0     \\
0       & 0     & 0     & 0             \\
\end{array}
\right) \, , \, 
\tilde C_8 = 
\frac{1}{\sqrt{3}}
\left(
\begin{array}{cccc}
0       & 0     & 0     & 0     \\
0       & - 2 i \sigma_3        & 0     & 0     \\
0       & 0     & i \sigma_3    & 0     \\
0       & 0     & 0     & i \sigma_3    \\
\end{array}
\right) 
\end{eqnarray}
Finally one can use the (unitary) permutation matrix
\begin{eqnarray}
T &=& 
\left(
\begin{array}{ccccccc}
1       & 0     & 0     & 0     & 0     & 0     & 0     \\
0       & 0     & 0     & 0     & 0     & 0     & 1     \\
0       & 0     & 0     & 0     & 1     & 0     & 0     \\
0       & 0     & 1     & 0     & 0     & 0     & 0     \\
0       & 0     & 0     & 0     & 0     & 1     & 0     \\
0       & 0     & 0     & 1     & 0     & 0     & 0     \\
0       & 1     & 0     & 0     & 0     & 0     & 0     
\label{eq:ad2} 
\end{array}
\right) 
\end{eqnarray}
to rearrange the rows and columns of $\tilde C_1 , \ldots , \tilde C_8$ 
such that it is obvious that 
the decomposition of the $7$-dimensional $SU(3)$ representation is 
given by $\{ 1 \} \oplus \{ 3 \}  \oplus \{ \bar 3 \}$.

\section{\label{sec:euler} Euler decomposition of G(2)} 

In order to implement a Metropolis update, 
we will use prototypes of $G(2)$ group elements to maneuver through 
group space. These are: 
\bea 
D_{1} (\alpha ) &=&  \exp \left\{ \alpha C_{1} \right\} 
\; = \; \left(
\begin{array}{ccccccc}
1    &   0   &   0   &   0   &   0   &   0   &   0 \\
0    &   1   &   0   &   0   &   0   &   0   &   0 \\
0    &   0   &   1   &   0   &   0   &   0   &   0 \\
0    &   0   &   0   &  \co  &   0   &   0   & -\si \\
0    &   0   &   0   &   0   &  \co  & -\si  &   0 \\
0    &   0   &   0   &   0   &  \si  &  \co  &   0 \\
0    &   0   &   0   &  \si  &   0  &   0   &  \co 
\end{array}
\right) ,
\non \\ 
D_{2} (\alpha ) &=& \exp \left\{ \alpha C_{2} \right\} 
\; = \; \left(
\begin{array}{ccccccc}
1    &   0   &   0   &   0   &   0   &   0   &   0 \\
0    &   1   &   0   &   0   &   0   &   0   &   0 \\
0    &   0   &   1   &   0   &   0   &   0   &   0 \\
0    &   0   &   0   &  \co  &   0   &  \si  &   0 \\
0    &   0   &   0   &   0   &  \co  &   0   & -\si \\
0    &   0   &   0   & -\si  &   0   &  \co  &   0 \\
0    &   0   &   0   &   0   &  \si  &   0   &  \co 
\end{array}
\right) , 
\non \\ 
D_{3} (\alpha ) &=& \exp \left\{ \alpha C_{3} \right\} 
\; = \; \left(
\begin{array}{ccccccc}
1    &   0   &   0   &   0   &   0   &   0   &   0 \\
0    &   1   &   0   &   0   &   0   &   0   &   0 \\
0    &   0   &   1   &   0   &   0   &   0   &   0 \\
0    &   0   &   0   &  \co  & -\si  &   0   &   0 \\
0    &   0   &   0   &  \si  &  \co  &   0   &   0 \\
0    &   0   &   0   &   0   &   0   &  \co  & -\si \\
0    &   0   &   0   &   0   &   0   &  \si  &  \co 
\end{array}
\right) , 
\non \\
D_{4} (\alpha ) &=& \exp \left\{ \alpha C_{4} \right\} 
\; = \; \left(
\begin{array}{ccccccc}
1    &   0   &   0   &   0   &   0   &   0   &   0 \\
0    &  \co  &   0   &   0   &   0   &   0   &  \si \\
0    &   0   &  \co  &   0   &   0   &  \si  &   0 \\
0    &   0   &   0   &   1   &   0   &   0   &   0 \\
0    &   0   &   0   &   0   &   1   &   0   &   0 \\
0    &   0   & -\si  &   0   &   0   &  \co  &   0 \\
0    & -\si  &   0   &   0   &   0   &   0   &  \co 
\end{array}
\right) , 
\non \\ 
D_{5} (\alpha ) &=& \exp \left\{ \alpha C_{5} \right\} 
\; = \; \left(
\begin{array}{ccccccc}
1    &   0   &   0   &   0   &   0   &   0   &   0 \\
0    &  \co  &   0   &   0   &   0   & -\si  &   0 \\
0    &   0   &  \co  &   0   &   0   &   0   &  \si \\
0    &   0   &   0   &   1   &   0   &   0   &   0 \\
0    &   0   &   0   &   0   &   1   &   0   &   0 \\
0    &  \si  &   0   &   0   &   0   &  \co  &   0 \\
0    &   0   & -\si  &   0   &   0   &   0   &  \co 
\end{array}
\right) , 
\non \\
D_{6} (\alpha ) &=& \exp \left\{ \alpha C_{6} \right\} 
\; = \; \left(
\begin{array}{ccccccc}
1    &   0   &   0   &   0   &   0   &   0   &   0 \\
0    &  \co  &   0   &   0   &  \si  &   0   &   0 \\
0    &   0   &  \co  & -\si  &   0   &   0   &   0 \\
0    &   0   &  \si  &  \co  &   0   &   0   &   0 \\
0    & -\si  &   0   &   0   &  \co  &   0   &   0 \\
0    &   0   &   0   &   0   &   0   &   1   &   0 \\
0    &   0   &   0   &   0   &   0   &   0   &   1 
\end{array}
\right) , 
\non \\ 
D_{7} (\alpha ) &=& \exp \left\{ \alpha  C_{7} \right\} 
\; = \; \left(
\begin{array}{ccccccc}
1    &   0   &   0   &   0   &   0   &   0   &   0 \\
0    &  \co  &   0   & -\si  &   0   &   0   &   0 \\
0    &   0   &  \co  &   0   & -\si  &   0   &   0 \\
0    &  \si  &   0   &  \co  &   0   &   0   &   0 \\
0    &   0   &  \si  &   0   &  \co  &   0   &   0 \\
0    &   0   &   0   &   0   &   0   &   1   &   0 \\
0    &   0   &   0   &   0   &   0   &   0   &   1 
\end{array}
\right)  
\eea
which are ``pure'' SU(2) subgroup elements. Furthermore we use: 
\bea
D_{8} (\alpha ) &=& \exp \left\{ \sqrt{3}\, \alpha C_{8} \right\} 
\; = \; \left(
\begin{array}{ccccccc}
1    &   0   &   0   &   0   &   0   &   0   &   0 \\
0    &  \cco & -\ssi &   0   &   0   &   0   &   0 \\
0    &  \ssi &  \cco &   0   &   0   &   0   &   0 \\
0    &   0   &   0   &  \co  &  \si  &   0   &   0 \\
0    &   0   &   0   &  -\si &  \co  &   0   &   0 \\
0    &   0   &   0   &   0   &   0   &  \co  &  -\si \\
0    &   0   &   0   &   0   &   0   &  \si  &  \co
\end{array}
\right) , 
\non \\ 
D_{9} (\alpha ) &=& \exp \left\{ \sqrt{3}\, \alpha C_{9} \right\} 
\; = \; \left(
\begin{array}{ccccccc}
\cco & -\ssi &   0   &   0   &   0   &   0   &   0 \\
\ssi &  \cco &   0   &   0   &   0   &   0   &   0 \\
0    &   0   &   1   &   0   &   0   &   0   &   0 \\
0    &   0   &   0   &  \co   &   0   &   0   &  \si \\
0    &   0   &   0   &   0   &  \co   & -\si   &   0 \\
0    &   0   &   0   &   0   &  \si   &  \co   &   0 \\
0    &   0   &   0   & -\si   &   0   &   0   &  \co 
\end{array}
\right) , 
\non \\ 
D_{10} (\alpha ) &=& \exp \left\{ \sqrt{3}\, \alpha C_{10} \right\} 
\; = \; \left(
\begin{array}{ccccccc}
\cco &   0   & -\ssi &   0   &   0   &   0   &   0 \\
0    &   1   &   0   &   0   &   0   &   0   &   0 \\
\ssi &   0   &  \cco &   0   &   0   &   0   &   0 \\
0    &   0   &   0   &  \co   &   0   & -\si   &   0 \\
0    &   0   &   0   &   0   &  \co   &   0   & -\si \\
0    &   0   &   0   &  \si   &   0   &  \co   &   0 \\
0    &   0   &   0   &   0   &  \si   &   0   &  \co 
\end{array}
\right) , 
\non \\
D_{11} (\alpha ) &=& \exp \left\{ \sqrt{3}\, \alpha C_{11} \right\} 
\; = \; \left(
\begin{array}{ccccccc}
\cco &   0   &   0   & -\ssi &   0   &   0   &   0 \\
0    &  \co  &   0   &   0   &   0   &   0   & -\si \\
0    &   0   &  \co  &   0   &   0   &  \si  &   0 \\
\ssi &   0   &   0   &  \cco &   0   &   0   &   0 \\
0    &   0   &   0   &   0   &   1   &   0   &   0 \\
0    &   0   & -\si  &   0   &   0   &  \co  &   0 \\
0    &  \si  &   0   &   0   &   0   &   0   &  \co 
\end{array}
\right) , 
\non \\ 
D_{12} (\alpha ) &=& \exp \left\{ \sqrt{3}\, \alpha C_{12} \right\} 
\; = \; \left(
\begin{array}{ccccccc}
\cco &   0   &   0   &   0   & -\ssi &   0   &   0 \\
0    &  \co  &   0   &   0   &   0   &  \si  &   0 \\
0    &   0   &  \co  &   0   &   0   &   0   &  \si \\
0    &   0   &   0   &   1   &   0   &   0   &   0 \\
\ssi &   0   &   0   &   0   &  \cco &   0   &   0 \\
0    & -\si  &   0   &   0   &   0   &  \co  &   0 \\
0    &   0   & -\si  &   0   &   0   &   0   &  \co 
\end{array}
\right) , 
\non \\ 
D_{13} (\alpha ) &=& \exp \left\{  \sqrt{3}\, \alpha C_{13} \right\} 
\; = \; \left(
\begin{array}{ccccccc}
\cco &   0   &   0   &   0   &   0   & -\ssi &   0 \\
0    &  \co  &   0   &   0   & -\si  &   0   &   0 \\
0    &   0   &  \co  & -\si  &   0   &   0   &   0 \\
0    &   0   &  \si  &  \co  &   0   &   0   &   0 \\
0    &  \si  &   0   &   0   &  \co  &   0   &   0 \\
\ssi &   0   &   0   &   0   &   0   &  \cco &   0 \\
0    &   0   &   0   &   0   &   0   &   0   &   1 
\end{array}
\right) , 
\non \\ 
D_{14} (\alpha ) &=& \exp \left\{ \sqrt{3}\, \alpha C_{14} \right\} 
\; = \; \left(
\begin{array}{ccccccc}
\cco &   0   &   0   &   0   &   0   &   0   & -\ssi \\
0    &  \co  &   0   &  \si  &   0   &   0   &   0 \\
0    &   0   &  \co  &   0   & -\si  &   0   &   0 \\
0    & -\si  &   0   &  \co  &   0   &   0   &   0 \\
0    &   0   &  \si  &   0   &  \co  &   0   &   0 \\
0    &   0   &   0   &   0   &   0   &   1   &   0 \\
\ssi &   0   &   0   &   0   &   0   &   0   &   \cco  
\end{array}
\right) 
\eea

Following~\cite{Cacciatori:2005yb}, the Euler angle representation 
of an arbitrary G(2) element $G$ is given by: 
\bea 
G &=& D_8(\alpha _1) \, D_9(\alpha _2) \, D_8 (\alpha _3) \; 
D_3 (\alpha _4) \, D_2 (\alpha _5 ) \, D_3 (\alpha _6) \; 
\label{eq:b1} \\ 
&& D_{11} (\alpha _7) \, D_5 (\alpha _8) \; 
D_3 (\alpha _9) \, D_2 (\alpha _{10}) \, D_3 (\alpha _{11}) \; 
D_8 (\alpha _{12}) \, D_9 (\alpha _{13}) \, D_8 (\alpha _{14}) \; , 
\nonumber 
\eea
where the range of parameters is found to be: 
\beq
a_1, \; a_4 , \;  a_{9} , \; a_{12}  \; \in \;  [0, 2\pi [ \; , 
\hbo 
a_2, \; a_5 , \;  a_{10} , \; a_{13}  \;\in \;  [0, \frac{\pi }{2} [ \; , 
\hbo 
a_3, \; a_6 , \;  a_{11} , \; a_{14}  \; \in \; [0, \pi [ \; . 
\label{eq:b2} \\ 
\eeq 
and
\beq 
a_8 \; \in \;  [0, \frac{\pi }{2} [ \; , \hbo 
a_7 \; \in \;  [0, \frac{2 }{3} a_8 [ \; . 
\label{eq:b3} \\ 
\eeq 
\end{widetext}

\section{ \label{sec:g2_sub} Polyakov line 
 and the Cartan subgroup  } 

\subsection{  Higgs field and unitary gauge }

As pointed out by Holland et al.\ \cite{Pepe1}, a fundamental Higgs field will break
$G(2)$ gauge symmetry to $SU(3)$. This can be easily anticipated resorting 
to the real representation of $G(2)$. Also for later use, we will here 
provide the sequence of $SO(2)$ rotations $D_k(\alpha)$ with which 
the $7$ component Higgs field can be rotated into the $1$-direction:
\begin{eqnarray}
\nonumber
\left(
\begin{array}{c}
\ast \\ \ast \\ \ast \\ \ast \\ \ast \\ \ast \\ \ast 
\end{array}
\right)
& \stackrel{D_{10}(\alpha_1)}{\longrightarrow} &
\left(
\begin{array}{c}
\ast \\ \ast \\ 0 \\ \ast \\ \ast \\ \ast \\ \ast 
\end{array}
\right)
\stackrel{D_{9}(\alpha_2)}{\longrightarrow}
\left(
\begin{array}{c}
\ast \\ 0 \\ 0 \\ \ast \\ \ast \\ \ast \\ \ast 
\end{array}
\right)
\\
& \stackrel{D_{8}(\alpha_3)D_{3}(\alpha_3)}{\longrightarrow} &
\left(
\begin{array}{c}
\ast \\ 0 \\ 0 \\ \ast \\ \ast \\ \ast \\ 0
\end{array}
\right)
\stackrel{D_{8}(\alpha_4)D_{3}(-\alpha_4)}{\longrightarrow} 
\left(
\begin{array}{c}
\ast \\ 0 \\ 0 \\ \ast \\ 0 \\ \ast \\ 0
\end{array}
\right) 
\nonumber \\
& \stackrel{D_{2}(\alpha_5)}{\longrightarrow} &
\left(
\begin{array}{c}
\ast \\ 0 \\ 0 \\ \ast \\ 0 \\ 0 \\ 0
\end{array}
\right) 
\stackrel{D_{11}(\alpha_6)}{\longrightarrow} 
\left(
\begin{array}{c}
\ast \\ 0 \\ 0 \\ 0 \\ 0 \\ 0 \\ 0
\end{array}
\right) \, .
\end{eqnarray}
Here, the symbol $\ast $ denotes an arbitrary real number. The
different angles $\alpha_j$ can be easily computed
step by step. To be precise, we here provide some details. 
Let us consider the vector $(x,y)^T$ the $y$ component of which we would like 
to rotate to zero. There are two possibilities for that: After the 
rotation of the vector, its $x$ component can be either positive 
or negative. Throughout this section, we adopt the ``minimal'' choice, 
which preserves the sign of the $x$ component: 
$$
\left(
\begin{array}{cc}
\cos \phi & - \sin \phi \\ 
\sin \phi & \cos \phi 
\end{array}
\right) \; 
\left(
\begin{array}{c}
x \\ y 
\end{array}
\right) \; = \; 
\left(
\begin{array}{c}
\mathrm{sign}(x) \; \sqrt{x^2 + y^2 } \\ 0 
\end{array}
\right)
$$
where 
$$ 
\cos \phi \; = \; \frac{ \vert x \vert }{  \sqrt{x^2 + y^2 } } \, , \hbo 
\sin \phi \; = \; \frac{ - \mathrm{sign}(x) \; y 
}{  \sqrt{x^2 + y^2 } } \, . 
$$

Now we will show that a non-zero Higgs field breaks the
symmetry down to $SU(3)$. To this end we have to calculate the
stabilizer of the vector $\phi_0 = (1,0,0,0,0,0,0)^T$, i.e.~we have to
determine the $G(2)$ group elements $g$ fulfilling 
\beq
g \phi_0 = \phi_0 \, .
\label{eq:stab}
\eeq
If we are only interested in continuous subgroups as stabilizer we only
need to analyze the neighborhood of the identity. 
Infinitesimally eq.~(\ref{eq:stab}) translates to
\begin{eqnarray}
0 = \sum_k \beta_k C_k \phi_0 = 
\frac{2}{\sqrt{3}} (0,\beta_9,\beta_{10},\beta_{11},
\beta_{12},\beta_{13},\beta_{14})^T\, ,
\end{eqnarray}
i.e.~$\beta_9=\beta_{10}=\beta_{11}=\beta_{12}=\beta_{13}=\beta_{14}=
0$. But this means that the only continuous subgroup of $G(2)$ leaving
$\phi_0$ invariant is generated by $C_k \, , \, k=1,\ldots,8$, i.e.~it
is the $SU(3)$ subgroup. With our considerations we cannot exclude 
other elements of $G(2)$ (forming a discrete subgroup) which fulfil
eq.~(\ref{eq:stab}).

\subsection{ The Cartan subgroup }

The Polyakov line $P$ homogeneously transforms under the gauge transformation 
$\Omega $. Let $H$ be an element of G(2) with 
\beq 
P \; = \; \Omega \, H \, \Omega ^T \; . 
\label{eq:e1} 
\eeq 
For an arbitrary element $P$ of G(2), we introduce a corresponding 
7-dimensional vector with the help of the constraint constants 
$T_{abc}$ (\ref{eq:constr2}) by 
\beq 
(\vec{p} )_a \; := T_{abc} \; P_{bc} \; . 
\label{eq:e2} 
\eeq 
Rewriting (\ref{eq:constr1}) as 
$$ 
T_{auv} \; \Omega_{ub} \; \Omega_{vc} \; = \; \Omega_{am} \; T_{mbc} \; , 
$$ 
we easily show that the vectors $\vec{p}$, $\vec{h}$, constructed from 
$P$ and $H$ respectively, are related by 
\beq 
\vec{p} \; = \; \Omega \; \vec{h} \; . 
\label{eq:e3} 
\eeq 
In the last subsection, we have already verified that any 
(real) 7-dimensional vector can be rotated to $\vec{e}_1 = (1,0,0,0,0,0,0)^T$ 
direction using $\Omega \in G(2)$ only. 

\vskip 0.3cm 
After a suitable choice of gauge, we now consider Polyakov lines 
$P \in G(2)$ which satisfy $\vec{p} \propto \vec{e}_1$. 
Using the constraint (\ref{eq:constr1}), we find: 
$$ 
T_{def} \; P_{da} \; P_{eb} \; = \; T_{abc} \; P_{fc} \; , 
$$
and in particular (after summing over $b=f$)
$$ 
P^T \; \vec{p} \; = \; \vec{p} \; . 
$$
Hence, all elements $G\in G(2)$ with $\vec{g} \propto \vec{e}_1$ constitute 
the subgroup of rotations which leave the vector $\vec{e}_1$ 
invariant. For $\Omega _3 \subset SU(3)$, where 
$SU(3)$ is the subgroup of G(2) spanned by the generators $C_1 \ldots C_8$, 
we find 
$$ 
\vec{e}_1 \; = \; \Omega _3 \; \vec{e}_1 \; , \hbo 
$$
implying that these rotations are identified with the $SU(3)\subset G(2)$ 
subgroup. 

\vskip 0.3cm 
Finally, there are elements $\Omega _3 \in SU(3) \subset G(2)$ which 
transform $P \in SU(3) $ to its center elements. These elements 
constitute the rank 2 Cartan subgroup of $G(2)$.

\section{\label{sec:complex} Complex Representation and Gauge-Fixing}

Matrix $\Z$ has the form
\beq
            {\cal Z} = \left( \begin{array}{ccc}
                          C & \m K  &  D^*\cr
                        -\m K^\dg   & \frac{1-x}{ 1+x} & - \m K^T  \cr
                          D &  \m K^* & C^*  \end{array} \right)
\eeq
where $x = \parallel K \parallel^2$, $\m = \sqrt{2}/(1+x)$, and $C,D$ are 
the $3\times 3$ matrices
\bea
           C &=& \frac{1}{\D}\left\{ {\mathbf 1}_3 - \frac{M }{\D(1+\D)} \right\}
\non \\
           D &=& - \frac{W}{ \D} - \frac{S}{\D^2}
\eea
with
\bea
           M &=& K K^\dg ~~,~~ S = K^* K^\dg 
\non \\
           W &=& \epsilon_{\a\b\g} K_\g ~~,~~ \D=\sqrt{1+x}
\eea
Each group element is specified by fourteen parameters: eight for the SU(3) matrix $U$, and another
six for the complex 3-vector $K$.
   
   Our Monte Carlo updates combine a Cabbibo-Marinari update of the link variables, using group elements
in the SU(3) subgroup of G(2) (i.e.\ the ${\cal U}$ matrices), followed by a gauge 
transformation by the ${\cal Z}$ matrices,
chosen at random from a lookup table.\footnote{The idea of combining SU(3) Cabbibo-Marinari
updates with random G(2) gauge transformations was suggested to us by M. Pepe.}   The lookup table
of several thousand ${\cal Z}$ matrices is generated stochastically, by choosing the real and imaginary 
parts of each component of the complex $K$ vector from a uniform distribution of random numbers 
in the range $[-1,1]$.  For each ${\cal Z}$ matrix entered into the table, one also enters its inverse,
${\cal Z}^{-1} = {\cal Z}^\dg$. To carry out the Cabbibo-Marinari updates, it is necessary
to generate a matrix
\beq
                     \left( \begin{array}{ccc}
                          A &   & \cr
                            & 1 &  \cr
                            &   &  A^*  \end{array} \right)
\eeq
where the $3\times 3$ $A$ matrix belongs to one of the three standard SU(2) subgroups of SU(3), and has non-zero
elements only in the $i,j$-th rows and columns . Let $\G_l$ be the link variable at link $l$, $F$ the
associated sum of staples, and $R=\G_l F$.  Also denote $i'=i+4,~j'=j+4$, and
introduce the $2\times 2$ matrices
\bea
       a &=& \left( \begin{array} {cc}
                A_{ii} & A_{ij} \cr
                A_{ji} & A_{jj} \end{array} \right)
\non \\
       r &=& \left( \begin{array} {cc}
                R_{ii} + R^*_{i'i'} & R_{ij} + R^*_{i'j'} \cr
                R_{ji} + R^*_{j'i'} & R_{jj} + R^*_{j'j'} \end{array} \right)
\non \\
       a &=& a_0 \Id_2 + i \sum_{n=1}^3 a_n \s_n
\non \\
       r &=& r_0 \Id_2 - i \sum_{n=1}^3 r_n \s_n
\eea
where the $a_n$'s are real, and the $r_n$'s are, in general, complex.  Then
\beq
      \frac{\b }{ 7} \mbox{ReTr}[{\cal U}R] = \frac{2\b }{ 7} \sum_{n=0}^3 a_n \mbox{Re}(r_n)
\eeq
From here, one generates $a_n$ by the usual heat bath. This is done, at each site, for
each of the three usual SU(2) subgroups of SU(3).  

    We have checked that the plot of plaquette energies vs.\ coupling $\b$, generated by this
procedure, agrees with the corresponding result arrived at by a more conventional Metropolis updating
method, described in section \ref{sec:static} above.  The plot also agrees with the curve displayed in Fig.\ 2
of ref.\ \cite{Pepe2}.

  To fix to maximal SU(3) gauge, a simple steepest-descent algorithm seems to
be adequate.  We begin with a number of ``simulated annealing"
sweeps at zero temperature, i.e.\ at each site a random
complex 3-vector $K$ is generated, and the corresponding $\Z$ is used as a
trial gauge transformation.
If $R_1$ is lowered, the change is accepted.
This is followed by steepest descent sweeps. Denoting the
components of the $K$ vector as $K_i = x_i + i x_{i+3}$, we numerically compute
the gradient, at any given site
\beq
        (\nabla R)_i = \frac{\partial R }{ \partial x_i} ~~~~~i=1-6
\eeq
Then we set
\beq
       x_i = -(\nabla R)_i \epsilon
\eeq
where $\epsilon$ is gradually reduced, as the iterations proceed, so as not
to overshoot the minimum too often.   Although conjugate gradient methods
are preferable to steepest descent, in this case the number of iterations
required to converge to a local minimum of $R_1$ are not excessive.  The main
cost in gauge fixing comes at the second step, i.e.\ fixing to maximal 
$Z_3$ gauge.
We have found that steepest descent is inadequate for maximizing $R_2$.
Instead, at each site, we use a standard (quasi-Newton) optimization routine to maximize
$R_2$ with respect to the eight Euler angles \cite{Byrd} specifying an SU(3) gauge transformation
at that site, and proceed to fix the gauge by ordinary relaxation. 

      In order to do the SU(3) projection after maximal $Z_{3}$ gauge fixing, it
is necessary to factor each
link variable, which is a $7\times 7$ G(2) matrix, into the product $\Z \U$.  This
is carried out as follows:
For a given G(2) matrix $\G$, define the matrix
\beq
        M(K) = \Z^\dg(K) \G
\eeq
Then there is some choice of $K$ such that $M(K)=\U$, and $M(K)$ has the block-diagonal
form shown in eq.\ \rf{U}.  In that case there are 30 matrix elements of $M(K)$ which
vanish, and one that equals unity, but there are only six independent degrees
of freedom in $K$.  It is sufficient to compute $K$ from
only six of the block-diagonal conditions.  We have chosen these, somewhat arbitrarily,
to be
\bea
        f_1(K) &=& \mbox{Re}[M_{44} - 1] = 0
\non \\
        f_2(K) &=& \mbox{Re}[M_{35}] = 0
\non \\
        f_3(K) &=& \mbox{Re}[M_{26}] = 0
\non \\
        f_4(K) &=&\mbox{Re}[M_{52}] = 0
\non \\
        f_5(K) &=& \mbox{Re}[M_{61}] = 0
\non \\
        f_6(K) &=& \mbox{Re}[M_{73}] = 0
\eea
Solving the set of equations $f_i=0$ determines $K$.  This set can be solved numerically;
it can be checked that the corresponding $\U=M(K)$ satisfies all of the remaining block-diagonal conditions,
and the $3\times 3$ $U$ matrix and its complex conjugate in the non-zero blocks are unitary.

   The final step, projection of the $3\times 3$ SU(3) matrix $U$ to its nearest center element,
is standard.  The prescription is to select the SU(3) center element $z \in Z_3$ such that 
$z\Id_3$ is closest to the $3\times3$ SU(3) matrix $U$; i.e.\ choose the $z$ which maximizes
$\mbox{Re}[z^*\mbox{Tr}[U]]$. 

%
%

\end{document}